\documentclass[manuscript]{emulateapj}
\usepackage{color}
\usepackage{amsmath,natbib,graphicx}
\usepackage{graphics,graphicx}
\usepackage{epsf}
\usepackage{epstopdf}
\input{epsf}
\usepackage{epsfig}
\usepackage{color}
\DeclareGraphicsExtensions{.jpg,.pdf,.png,.eps,.ps}
\usepackage{hyperref}
\usepackage{ulem}
\usepackage{wrapfig}
\hypersetup{
    bookmarks=true,         % show bookmarks bar?
    unicode=false,          % non-Latin characters in Acrobat’s bookmarks
    pdftoolbar=true,        % show Acrobat’s toolbar?
    pdfmenubar=true,        % show Acrobat’s menu?
    pdffitwindow=false,     % window fit to page when opened
    pdfstartview={FitH},    % fits the width of the page to the window
    pdftitle={My title},    % title
    pdfauthor={Author},     % author
    pdfsubject={Subject},   % subject of the document
    pdfcreator={Creator},   % creator of the document
    pdfproducer={Producer}, % producer of the document
    pdfkeywords={keyword1} {key2} {key3}, % list of keywords
    pdfnewwindow=true,      % links in new window
    colorlinks=true,       % false: boxed links; true: colored links
    linkcolor=red,          % color of internal links
    citecolor=blue,        % color of links to bibliography
    filecolor=magenta,      % color of file links
    urlcolor=cyan           % color of external links
}

\def\red#1 {\textcolor{red}{#1}\ }   
\def\green#1 {\textcolor{green}{#1}\ }   
\def\cyan#1 {\textcolor{cyan}{#1}\ }

\def\gs{\mathrel{\raise0.35ex\hbox{$\scriptstyle >$}\kern-0.6em \lower0.40ex\hbox{{$\scriptstyle \sim$}}}}
\def\ls{\mathrel{\raise0.35ex\hbox{$\scriptstyle <$}\kern-0.6em \lower0.40ex\hbox{{$\scriptstyle \sim$}}}}

\shorttitle{The Global Star Formation Laws of Galaxies from a Radio Continuum Perspective}
\shortauthors{Liu et al.}

\begin{document}

%% LaTeX will automatically break titles if they run longer than
%% one line. However, you may use \\ to force a line break if
%% you desire.

\title{The Global Star Formation Laws of Galaxies \\
 from a Radio Continuum Perspective}

%% Use \author, \affil, and the \and command to format
%% author and affiliation information.
%% Note that \email has replaced the old \authoremail command
%% from AASTeX v4.0. You can use \email to mark an email address
%% anywhere in the paper, not just in the front matter.
%% As in the title, use \\ to force line breaks.

\author{
 L.~Liu\altaffilmark{1,2,4}, 
 Y.~Gao\altaffilmark{1},
 T.R.~Greve\altaffilmark{3},	
}
\altaffiltext{1}{Purple Mountain Observatory, Key Lab of Radio Astronomy, 2 West Beijing Road, 210008 Nanjing, PR China}
\altaffiltext{2}{Max-Planck-Institut f\"ur Radioastronomie, Auf dem H\"ugel 69, D-53121 Bonn, Germany}
\altaffiltext{3}{Department of Physics and Astronomy, University College London, Gower Street, London WC1E 6BT, UK}
\altaffiltext{4}{University of Chinese Academy of Sciences, 19A Yuquan Road, PO Box 3908, 100039 Beijing, PR China}
\email{ljliu@mpifri-bonn.mpg.de, yugao@pmo.ac.cn}

%================================
% ABSTRACT
%================================
\begin{abstract}
We study the global star formation law - the relation between gas and star formation (SF)
rates in a sample of 181 local galaxies with infrared (IR) luminosities spanning 
almost five orders of magnitude (${\rm 10^{7.8}-10^{12.3} ~ L_\odot}$),
which includes 115 normal spiral galaxies and 66 (ultra)luminous IR 
galaxies [(U)LIRGs, $L_{\rm IR}~\geq~10^{11}~L_\odot$].
We derive their atomic, molecular gas and dense molecular gas masses using 
newly available HI, CO and HCN data from the literature,
and SF rates are determined both from total IR (${\rm 8-1000~\mu m}$) and 
1.4 GHz radio continuum (RC) luminosities. 
In order to derive the disk-averaged surface densities of gas and SF rates,
we have taken a novel approach and used high-resolution RC
observations to measure the radio sizes for all 181 galaxies.
In our sample, we find that the surface density of dense molecular gas (as traced by HCN)
has the tightest correlation with that of SF rates ($\Sigma _ {\rm {SFR}}$),
and is linear in $\log-\log$ space (power-law slope of $N=1.01 \pm
0.02$) across the full galaxy sample. 
The correlation between surface densities of molecular gas ($\Sigma _ {\rm H_2}$,traced by CO) 
and $\Sigma _ {\rm {SFR}}$ is sensitive to the adopted value of the CO-to-${\rm H_2}$ 
conversion factor ($\alpha_{\rm CO}$) used to infer molecular gas masses from CO luminosities.
For a fixed Galactic value of $\alpha_{\rm CO}$, a power law index of $1.14 \pm 0.02$ is found.
If instead we adopt values for $\alpha_{\rm CO}$ of 4.6 and 0.8 for disk galaxies and (U)LIRGs,
respectively, we find the two galaxy populations separate into two distinct $\Sigma _ {\rm {SFR}}$
versus $\Sigma _ {\rm H_2}$ relations.
Finally, applying a continuously varying $\alpha_{\rm CO}$ to our sample,
we recover a single ${\rm \Sigma_{SFR}}$-$\Sigma_{\rm H_2}$ relation with 
slope of $1.60 \pm 0.03$.
The ${\rm \Sigma_{SFR}}$ is a steeper function of total gas $\Sigma_{\rm gas}$ (molecular gas with 
atomic gas) than that of molecular gas  $\Sigma_{\rm H_2}$,
and are tighter among low-luminosity galaxies.
We find no correlation between global surface densities of SFRs and
atomic gas (H\,{\scriptsize I}).

\end{abstract}

% Keywords should appear after the \end{abstract} command. The uncommented
%% example has been keyed in ApJ style. See the instructions to authors
%% for the journal to which you are submitting your paper to determine
%% what keyword punctuation is appropriate.


%=================================
%INTRODUCTION
%=================================

\section{Introduction}
The notion of a general so-called star formation (SF) law applicable to galaxies
was first introduced by \citet{schmidt1959}, who suggested
that the star formation rate (SFR) volume density ($\rho_{\rm
{SFR}}$) varies with the interstellar gas volume density ($\rho_{\rm {gas}}$) as
a power law, i.e.\ $\rho _ {\rm {SFR}} \propto \rho _ {\rm {gas}} ^ {N}$, where
$N \sim 2$.  Under the assumption of a constant scale-height for the star
forming gas, this relation translates directly into an equivalent expression for
the SFR and gas surface density,  $\Sigma _ {\rm {SFR}} \propto \Sigma
_ {\rm {gas}} ^ {N}$, which can be determined from observations.  In a seminal
study of the SFR (traced by H$\alpha$) and gas (traced by
H\,{\scriptsize I} and CO) distributions in a sample of nearby galaxies,
\citet{kennicutt1998} (K98 hereafter) found $N \sim 1.4 \pm 0.2$.  Nevertheless,
discrepancies widely exist in the power-law index
\citep[e.g.,][]{bouche2007,bigiel2008,blanc2009,verley2010}.  The exact
functional form of the SF law has important implications for our understanding
of the physical processes that govern SF in galaxies \citep{krumholz2012} and
also plays an important role in the evolution of global galaxy properties
\citep{feldmann2013}

In recent years, the increases in receiver sensitivities have allowed for
high-resolution maps of the SFR and gas distribution, and thus made
probing the spatially resolved SF law on sub-kpc scales in nearby galaxies
achievable. \citep[e.g.,][]{wong2002,kennicutt2007,bigiel2008,bigiel2011,
blanc2009,schruba2010,schruba2011,rahman2012}.  Such detailed studies provide
important insights into the physical processes governing the SF.  However, they
are rarely feasible in the distant Universe, and have only been achieved in a
few local (ultra)luminous IR galaxies [(U)LIRGs, $L_{\rm
IR}~\geq~10^{11}~L_\odot$] \citep[e.g.,][]{boquien2011} and high-$z$ galaxies
\citep[e.g.,][]{hodge2012}.  Instead, the global galaxy-averaged measurements,
covering a large range of galaxies with extremely distributed SFRs and gas
densities, provide a powerful way of characterising the SF law as a function of
galaxy-wide properties.

A critical aspect of determining the global SF law is the ability to accurately
measure the physical size of the star forming region within galaxies.  Past
studies have used a variety of optical, IR and CO imaging to determine the
extent of star forming areas.  In K98, where a total of more than 90 local
galaxies were studied (most are normal spiral galaxies, with  $\sim 10$
(U)LIRGs), the sizes used to normalize the total SFRs and gas masses for the
subset of normal galaxies were determined from optical images, while the sizes
of IR starbursts were determined from CO or IR maps.  Determining the physical
sizes of star forming regions in high-$z$ galaxies turned out to be a much
bigger challenge due to its requirement for higher resolution and sensitivity.
As a result, the high-$z$ studies, as well as local studies, usually adopted
different methods in different wavebands in deducing the sizes for the various
sub-samples. This would inevitably introduce systematic uncertainties.  Thus, it
is necessary to introduce a uniform way to determine the physical scales of both
local and high-$z$ galaxies.

In the work presented here, we revisit the local SF law based on a dataset
nearly twice as large as that of K98, and we use high-resolution radio continuum
(RC) maps exclusively to infer the sizes of star forming areas and their SFRs.
The radio is an excellent tracer of SF. It is extinction-free and exhibits an
extraordinarily tight correlation with far-IR light ($\sim 0.3$ dex scatter over
five orders of magnitude in luminosity) \citep{condon1991,yun2001,murphy2006b},
which shows little or no evolution over the redshift range $z=0-2$
\citep{sargent2010,bourne2011,mao2011}.  Spatially resolved measurements of
local star forming galaxies in the radio and far-IR have shown that this tight
relationship extends down to $\ls 1\,{\rm kpc}$ within disks
\citep{murphy2006a,murphy2006b,murphy2008}. Similar observations of early-type
galaxies show a striking agreement between the radio, mid-IR (${\rm 24 \mu m}$)
and CO morphologies \citep{lucero2007,young2009}.  The highest resolution of
radio interferometers (e.g., (J)VLA, eMerlin) can resolve even the most compact
nearby starburst galaxies
\citep{condon1996,carilli1998,carilli2000,pihlstrom2001,momjian2003}.  We
therefore expect our radio-inferred galaxy sizes to accurately match the star
forming regions in various types of galaxies in our sample.

Millimeter and IR observations of giant molecular clouds (GMCs) in our Galaxy
have shown that stars form in dense cores, and that massive stars form almost
exclusively in clumps/clusters of massive dense cores
\citep{evans1999,evans2008,wu2010}.  The dense gas ($n > 10^4~{\rm cm}^{-3}$)
residing in these cores is best traced by molecules with high-dipole moments and
thus high critical densities, such as HCN and CS.  \citet{gao2004a,gao2004b}
(GS04a,b hereafter) carried out the first systematic HCN survey of a large local
galaxy sample and found a tight linear correlation between the IR and HCN
luminosities spanning three orders of magnitude.  More recent studies, targeting
not just HCN but also other high-density gas tracers (e.g.  {\rm HCO$^{+}$},
HNC, CN and CS) in local galaxies, also find molecular line luminosities
linearly increasing with IR luminosity (${\rm L_{IR}}$)
\citep{baan2008,krips2008,juneau2009, matsushita2010,zhang2014}.  Here, we also
present an analysis of the dense gas SF law based on HCN observations, in
addition to the  SF law of atomic, molecular and total gas.
In order to obtain the largest possible HCN sample, we compiled all well-sampled
HCN measurements in local galaxies available in the literature.  

In the work presented here, we scale the various gas components (atomic,
molecular, total gas and dense molecular gas) and uniformly calibrated SFRs
(either IR or RC) with the consistently derived radio sizes to derive the
surface densities and their relations.  A brief report of some preliminary
results based on a smaller sample can be found in an earlier paper
\citep{liu2012}.  Here, we further increase our sample size by nearly 50\% and
present a thorough analysis and discussion.  We describe our galaxy sample and
data in \S\ \ref{section:sample and data}, and describe the method to measure
galaxy sizes from radio maps in \S\ \ref{section:radio size and SFR}.  Our derivation and analysis  of the global SF law in its
various forms are presented in \S\ \ref{section:results}.  We discuss various
related issues and factors that may have influence to the derived SF laws in \S\
\ref{section:discussion}, and conclude in \S\ \ref{section:summary}.  Throughout
this paper, we adopt a flat cosmology with  $H_{\rm 0}~=~71~{\rm
km~s^{-1}~Mpc^{-1}}$ \citep{komatsu2011}.

%=================================
%Sample and Data
%=================================

\section{Galaxy Sample and Data} \label{section:sample and data}
%----------------------------
%2.1 Sample
%----------------------------
\subsection{Sample selection} 
Our sample is a combined sample of K98 and GS04b which contains 130 galaxies,
supplemented by 51 additional sources with well-sampled HCN measurements
available in the literature.  K98 includes 61 spiral disk galaxies and 36
IR-selected  starburst galaxies (circumnuclear IR
starbursts in normal star forming galaxies and (U)LIRGs).  GS04b contains 65
galaxies, of which $\sim 25$ overlap with KS98 sample.  Finally, we excluded
galaxies with known powerful active galactic nuclei (AGN) and galaxies with
$q$-indices \citep{helou1985} less than 1.7.  The final sample comprises 181
local galaxies, 66 of which are (U)LIRGs.  It covers almost all morphological
types, ranging from early types (E, S0) to late spirals and irregular galaxies.
They are in a variety of environments, from apparently isolated galaxy to group
or cluster members.  The total-IR (8 -- 1000 ${\rm \mu m}$) luminosity of our
galaxies spans almost 5 orders of magnitude, ${\rm log}(L_{\rm IR}/{\rm
L_\odot})~\sim~7.8-12.3$.  The distances range from 0.75 to 280 Mpc.  For each
galaxy in the sample, we have high-resolution RC maps and integrated
H\,{\scriptsize I}/CO/IR/RC fluxes.  Of the entire sample of 181 galaxies, 128
galaxies have published HCN data in the literature, and 95 galaxies have
published H\,{\scriptsize I} maps or H\,{\scriptsize I} sizes.  Our sample is
the largest sample of local star forming galaxies with H\,{\scriptsize
I}/CO/HCN/IR/RC measurements available.  All luminosities ($L_{\rm
H\,{\scriptsize I}}$, $L_{\rm CO}$, $L_{\rm HCN}$, $L_{\rm IR}$ and $L_{\rm
RC}$) and luminosity distances were converted to the common cosmological
parameters adopted here.

%---------------------------
%2.2 Atomic Mass
%----------------------------
\subsection{Data}
\subsubsection{H\,{\scriptsize I} data}\label{subsubsection:HI-data}
H\,{\scriptsize I} maps or sizes for 95 of the 115 normal star forming galaxies
in our sample were obtained from H\,{\scriptsize I} imaging surveys carried out
with the (J)VLA and the Westerbork Synthesis Radio Telescope (WSRT).  They are:
(1) the H\,{\scriptsize I} Nearby Galaxy Survey (TH\,{\scriptsize I}NGS)
\citep{walter2008}, which provides high-resolution ($FWHM \sim 6 \arcsec$) VLA
maps of nearby galaxies; (2) the VLA Imaging survey of Virgo galaxies in Atomic
gas (VIVA) \citep{chung-A2009}, which includes H\,{\scriptsize I} images
($FWHM\sim 15\arcsec -16 \arcsec$) of late type galaxies in the Virgo cluster;
(3) Local Irregulars That Trace Luminosity Extremes, The H\,{\scriptsize I}
Nearby Galaxy Survey (Little TH\,{\scriptsize I}NGS) \citep{hunter2012}, which
assembles a dataset  of high sensitivity and resolution ($FWHM
\sim 6 \arcsec$) H\,{\scriptsize I} maps of dwarf irregular galaxies obtained
with the (J)VLA; (4) the Westerbork H\,{\scriptsize I} Survey of Spiral and
Irregular Galaxies  (WH\,{\scriptsize I}SP)
\citep{swaters2002a,swaters2002b,noordermeer2005}, a large H\,{\scriptsize I}
survey of several hundreds of nearby spiral and irregular galaxies ($FWHM\sim
15\arcsec \times 30 \arcsec$); (5) the Westerbork SAURON H\,{\scriptsize I}
Survey \citep{oosterloo2010}, which presents deep H\,{\scriptsize I} maps for
SAURON early-type galaxies ($FWHM\sim 70\arcsec \times 25 \arcsec$). 

The H\,{\scriptsize I} surface densities ($\Sigma_{\rm H\,{\scriptsize I}}$)
for these 95 galaxies were derived in one of the following two ways.  
For 51 galaxies with H\,{\scriptsize I} maps publically available as {\tt fits}
files, we directly measured the average H\,{\scriptsize I} surface densities
within our derived radio sizes (see \S\ \ref{section:radio size and SFR}
for how we derived the radio sizes).  For 44 of the galaxies, only
published H\,{\scriptsize I} disk sizes (defined as where the face-on
corrected H\,{\scriptsize I} surface density drops below $1~{\rm
M_{\odot}~pc^{-2}}$) were available, we averaged the total
H\,{\scriptsize I} mass over the H\,{\scriptsize I} sizes.  
We see no systematic differences in the H\,{\scriptsize I} surface densities between
the two cases (see Fig. \ref{figure:HI_IR}).
Comparing the H\,{\scriptsize I} sizes with the radio
sizes (see \S\ \ref{section:radio size and SFR}) we found the former to
be $3-7\times$ larger, which is not surprising since H\,{\scriptsize I} is often
found to extend to the optical diameter of disk galaxies or even to larger scale lengths 
where star formation is not occuring \citep{garciaruiz2002,swaters2002a,chung-A2009}.
For the remaining 20 normal star forming galaxies no published
H\,{\scriptsize I} maps or sizes were available, and we do not include them in
our analysis of the $\Sigma_{\rm SFR}$-$\Sigma_{\rm H\,{\scriptsize I}}$ relation.

We ignored the neutral atomic hydrogen in (U)LIRGs for the following reasons.
First, the atomic gas fraction in (U)LIRGs is much smaller, and often
negligible, compared with the molecular gas content, particularly within the
compact starburst regions.  It is well known that the ${\rm
H_2}$/H\,{\scriptsize I} ratio increases with the IR excess in (U)LIRGs
\citep{mirabel1989} and active star forming spiral galaxies tend to have more
than one order of magnitude higher ${\rm H_2}$/H\,{\scriptsize I} ratio than
those of the quiescent galaxies of extremely late Hubble types at lower
luminosities \citep{young1989}.  Secondly, the H\,{\scriptsize I} distribution
in (U)LIRGs often show complex morphologies, being distorted as the merger
progresses \citep{hibbard2001b,wang2001,chen2002}, or revealing elongated
streams or tails of gas \citep[e.g.,][]{taramopoulos2001,hibbard2001a}.  This
also tends to clearly show some signatures in the global H\,{\scriptsize I}
spectra in ULIRGs \citep[e.g.,][]{vandriel2001}.  In addition, strong
H\,{\scriptsize I} absorption feature is often seen against the nuclear regions
of (U)LIRGs \citep{cole1999,taramopoulos2001,chen2002,momjian2003}.

%---------------------------
%2.3 Molecular Mass
%----------------------------
\subsubsection{CO data}
Velocity-integrated CO fluxes were gathered from the literature
and, when available, preference was given to CO fluxes 
derived directly from galaxy-integrated imaging measurements,
of which we chose the values from single-dish maps 
\citep[GS04b;][]{kuno2007,chung-E2009} over interferometer maps 
\citep{sofue2003,helfer2003,young2008} to avoid missing flux. 
Our second choice was the global CO fluxes inferred from observations along the
major and/or minor axes of the disks, combining with model assumptions of the gas
distribution \citep[GS04b;][]{young1995,paglione2001,nishiyama2001}.
Instead of early CO surveys compiled in K98's work, which suffered from 
both small sample sizes and limited spatial coverage as well as sensitivity, 
the new CO surveys had made impressive progress 
in both sampled spatial coverage and observation resolution.  
The improved sensitivity of these observation and facilities even allowed 
the cold molecular gas of early-type galaxies to be detected  
\citep[e.g.,][]{young2008,crocker2011}.
We found a general agreement between the CO fluxes from the different surveys 
in the cases of overlapping CO measurements. The agreement is typically  
within 20\% - 35\%, although some interferometric fluxes deviated by as much as 
20\% - 50\% from the single dish measurements. 
The data sets from surveys along major and/or minor axis deviated from 
single dish fluxes by smaller factors of 5\% - 30\%.

%---------------------------
%2.4 Dense Molecular Mass
%----------------------------
\subsubsection{HCN data}
We used the HCN (1-0) transition 
($n_{\rm crit} \approx 1 \times 10^5 ~{\rm cm^{-3}}$) 
as a tracer of dense molecular gas,
as it is one of the most abundant high dipole moment molecules
and the most frequently observed interstellar molecules after CO in galaxies.
In order to obtain the largest possible HCN (1-0) sample, 
we collected all robust measurements available in the literature.
Our final HCN sample consists of 128 galaxies in total, mostly from GS04b, 
\citet{baan2008}, and \citet{GB-2012}.
It also includes recent HCN (1-0) detections of a handful of CO-luminous early-type galaxies 
\citep{krips2010,crocker2012}, which enabled us to probe the dense gas SF law
at lower luminosities. 
The compiled sample spanned 3 orders in HCN luminosity 
($L'_{\rm HCN}/ ({\rm K~km~s^{-1}~pc^2})~\sim~10^{7.1-9.7}$).
It consists of 65 (U)LIRGs, 47 spiral galaxies, and 16 early-type galaxies.
The formulas to calculate atomic gas mass, molecular mass,
and dense molecular mass from the observed line fluxes are given in Appendix \ref{appendix:gas mass and SFR}.

\subsubsection{IR and 1.4 GHz RC data}\label{subsubsection:IR-radio-data}
The IR emission is a sensitive tracer of the young stellar population and SFR,
as a significant fraction of the bolometric luminosity of a galaxy is absorbed
by interstellar dust and re-emitted in the thermal IR (K98).  Total-IR 8 -- 1000
${\rm \mu m}$ luminosities were calculated using 12, 25, 60, 100 ${\rm \mu m}$
photometry from the {\sl Infrared Astronomical Satellite (IRAS)}  and the formulas presented in Table 1 of \citet{sanders1996}.

The 1.4 GHz radio continuum emission probes synchrotron radiation 
from supernovae remnants and thermal emission from H\,{\scriptsize II} regions, 
whose progenitors are both short-lived massive stars and 
hence trace recent SF \citep[e.g.,][]{condon1992,yun2001}. 
This make the radio a reliable tracer of SF,
as indicated by the tight IR-radio correlation.
Integrated 1.4 GHz radio fluxes were, for the most part, 
obtained from the NRAO VLA Sky Survey \citep[NVSS;][]{condon1998,yun2001,condon2002}.
Due to its large beam size and high sensitivity,  
the NVSS is able to provide reliable integrated flux measurements.
For the remaining sources, 1.4 GHz fluxes were obtained from a number of other radio surveys,
see Table\ \ref{table:table1} for references. Mono-chromatic (1.4 GHz) radio luminosities
were subsequently inferred using $L_{\rm 1.4GHz} = 4\pi D_{\rm L}^2 S_{\rm 1.4GHz}$. 
Table\ \ref{table:table1} lists the distance, as well as CO, HCN, IR and RC luminosities.

%%%%%%%%%%%%%%%%%%
%Sizes and SFRs estimates
%%%%%%%%%%%%%%%%%%
\section{Radio sizes and star formation rates} \label{section:radio size and SFR}
We do not only use integrated radio (1.4 GHz) fluxes to trace SFRs of galaxies,
but also utilize radio images to derive the sizes of star forming regions.
About two thirds of our sources  had published
high-resolution radio maps from various systematic VLA radio imaging surveys
\citep[]{condon1987,condon1990,condon1991,condon1996,becker2003}.   The  remaining one
third ($\sim64$ sources)  did not however,  (or in some cases they were of poor quality), 
and in those cases we had to reduce the  raw
data retrieved from the NRAO archive (https://archive.nrao.edu) using
 $\cal AIPS\/$.  The image reduction was done using standard
 $\cal AIPS\/$ procedures, i.e.\ blanking of bad visibilities,
followed by calibration and CLEAN imaging.  In some cases, data sets from
different VLA configurations were combined in the visibility plane to improve
sensitivity and resolution.  This step was done using the task DBCON after each
$uv$ data set had been calibrated separately.  For a few ($\sim3$) southern
sources, radio maps from the Molonglo Observatory Synthesis Telescope (MOST)
were used \citep{mauch2008}.  As a consistency check, we compared the NVSS radio
fluxes with the integrated fluxes derived from our reduced maps and found good
agreement.

The typical noise level of our radio maps is ${\rm \sim0.1~mJy~beam^{-1}}$,  
which should be sufficient to detect low-brightness emission and 
allow for accurate measurement of the radio size.
Radio sizes were determined from 2D Gaussian fits in the image plane
using the  $\cal AIPS\/$ task IMFIT. 
A detailed description of the radio size measurements and related issues,
such as AGN contamination and missing flux problem due to high-resolution interferometer data, 
are given in the Appendix \ref{appendix:radio size measurement}.

The radio size may arguably serve as a more truthful picture of the extent of the star forming 
regions than the mixture of IR and CO maps often used, since the radio continuum is 
a well-known, extinction-free SF tracer \citep{condon1991,condon1996} while
both of IR and CO observations often suffer from very limited sensitivity and resolution. 
Furthermore, in some cases CO may also be tracing diffuse molecular gas 
that is not actively forming stars \citep{bothwell2010,rujopakarn2011}, 
while the IR emission have a significant contribution from dust
heated by the ambient FUV field from evolved stellar population. 
The radio continuum, on the other hand, probes synchrotron radiation from 
supernovae remnants and thermal emission from H\,{\scriptsize II}
regions, both of which have short-lived massive stars as progenitors, and may therefore be a better
tracer of the recent SF and its spatial extent within galaxies.  

Sizes inferred from optical images, often adopted for normal galaxies, include old
stellar components that usually extend beyond the regions where the SF is occurring.
Indeed, a comparison of the optical sizes of the normal galaxies from K98 with our measured
radio sizes shows the former to be $2-5\times$ larger, with an average value of $\sim 3.3$.
The approach of utilizing radio continuum maps provides a more 
uniform way to determine the SFRs as well as the sizes of the 
actively star forming regions for a variety galaxy types.

The ideal way to derive disk-averaged surface densities of molecular and dense 
molecular gas is to measure them directly from CO and HCN maps.
Unfortunately, this is impractical for large sample of galaxies at the moment
owing to the limited sensitivity and weakness of HCN emission.
The strategy adopted here is to use our measured RC sizes as 
an estimate of the spatial extent of the molecular gas as well. 
This is a reasonable assumption since CO and HCN are known to trace SF in our 
own Galaxy \citep{wu2005} and in external galaxies \citep{murgia2002,iono2009}, 
and so should be tightly correlated with RC which is itself a tracer 
of ongoing formation of massive stars \citep{liu2010}.
Indeed, morphological similarities between RC and the molecular gas has not only been 
found in normal spiral galaxies \citep{crosthwaite2002,schuster2007} and (U)LIRGs 
\citep{condon1991,downes1998,murgia2005,tacconi2006}, 
but also in  early type galaxies \citep{lucero2007}.

\begin{figure}[h]
\includegraphics[scale=.50]{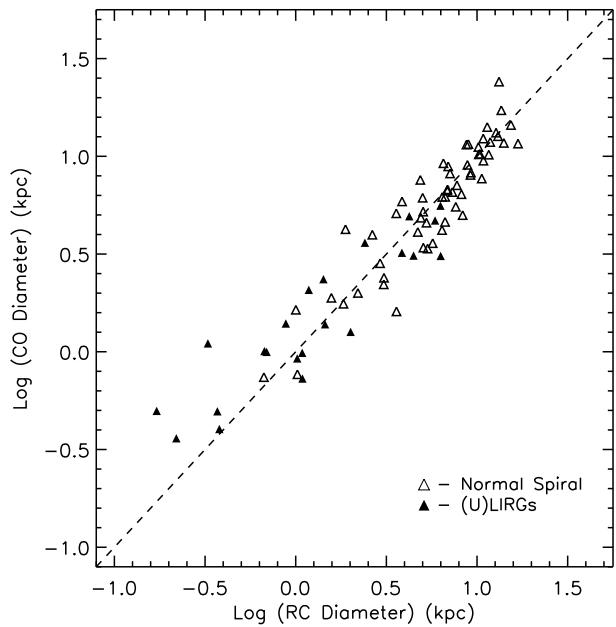}
\caption{CO size vs. RC size for 81 galaxies in our sample.
The distribution of CO and RC sizes generally scale linearly (dashed line).
This indicates that the RC and CO emission generally trace each other well.
The sizes of normal disk galaxies span a range of ${\rm \sim 1-15~kpc}$,
while ULIRGs are much more compact with size ${\rm \le 1.5~kpc}$.
At the low end, the CO sizes seem to be systematically larger than 
the RC sizes, suggesting CO could also trace diffuse molecular gas 
that is not actively forming stars.
}
\label{figure:co(size)-RC(size)}
\end{figure}

Comparing the radio sizes of a subsample of 81 galaxies with their CO sizes
collected from the literature, a quite good agreement was found 
(Fig.\ \ref{figure:co(size)-RC(size)}).  
CO sizes differ from RC sizes within a factor $\sim 2-3$, with an average ratio of $\sim 1.1$.
The CO sizes were derived from: (1) interferometry measurements
\citep[e.g.,][]{bryant1999,regan2001,iono2009}; (2) single-dish maps
\citep[e.g.,][]{leroy2009,chung-E2009}; (3) the observations along the major
and/or minor axes of the disks \citep[e.g.,][]{young1995} in literature.  The
good correlation between RC size and CO size indicates that the  RC  traces the 
global CO emission reasonably well.  However, we  notice that the CO
sizes seem to be systematically larger than the RC sizes at the low end 
(Fig.\ \ref{figure:co(size)-RC(size)}) which could be due to  CO also  tracing 
diffuse non-SF gas  as suggested above.  In extremely compact starbursts, the typical size
scale of SF is much smaller than $\sim 1$ kpc.

%==========================
%Figure showing CO versus RC sizes
%==========================

The SFRs were inferred in two independent ways, namely from the IR luminosities
and the rest-frame 1.4 GHz luminosities derived in
\S~\ref{subsubsection:IR-radio-data}.  To this end we adopted the
 luminosity-SFR calibrations provided by
\citet{bell2003} which account for the reduced efficiency of IR and radio
emission in low-luminosity galaxies (i.e., the IR and radio emission underestimate
the SFRs of low-luminosity galaxies) \citep{price1992,wang1996}. 
We describe the detailed formulas for calculating SFRs in Appendix
\ref{appendix:SFRs}.  Overall, the IR- and RC-based SFRs were
 in agreement ( see Appendix \ref{appendix:SFRs} for details) and we
shall use the IR-based values in the following.
But some examples of SF laws derived by adopting RC-based SFRs 
were given in Appendix \ref{appendix:SFRs}.  The resulting
deconvolved major axes, as well as surface densities for atomic gas, molecular
gas, total gas, dense molecular gas and IR/RC-based SFRs are 
given in Table\ \ref{table:table2}.

%%%%%%%%%%%%%%%%%%
%  Result
%%%%%%%%%%%%%%%%%%
\section{Results} \label{section:results}
The surface densities of molecular, total gas, dense molecular gas and SFRs are
derived by the formula $\Sigma =  \frac{\rm {Integrated~Value}}{2 \pi r^2_{\rm
RC}}$, where ${r_{\rm RC}}$ is the deconvolved HWHM along the major axis of the
RC maps (see Table\ \ref{table:table2}) to account for the galaxy inclination
 to first order.  The factor 1/2 is introduced to provide a
better estimate of observed surface densities over galaxies.

A realistic estimate of the uncertainty on the observed CO/HCN/IR/RC fluxes is $\sim 30\%$.  
However, considering the additional uncertainties in the calibrations that convert 
observed fluxes to physical quantities (i.e., luminosities, SFRs, and gas masses) as well as
measurement errors in radio sizes, we adopted a systematic uncertainty of 
$50\%$ (i.e., $0.22\,{\rm dex}$) on the final surface densities of SFR and gas.  
To compute the linear regression between surface densities of SFRs and gas we used
the publicly available Bayesian Markov Chain Monte Carlo (MCMC) IDL routine {\tt
linmix\_err} \citep{kelly2007}, which accounts for measurement errors in the x
and y variables, upper limits and intrinsic scatters.

We fit our SF laws with three parameters: power-law index N, normalization A
and intrinsic scatter $\epsilon$.  The functional of form is: $\log \Sigma_{\rm
SFR} = N\log \Sigma_{\rm gas} + A + \mathcal{N}(0,\epsilon)$, where
$\mathcal{N}(0,\epsilon)$ is introduced as a logarithmic deviation from the
power law, drawn from a normal distribution with zero mean and standard
deviation $\epsilon$.  The intrinsic scatter $\epsilon$ is caused by
genuine differences in the SF properties of our galaxies, yet the  conservative systematic
errors ($50\%$) adopted prevent us from extracting this  information
(i.e., the fit returns $\epsilon \sim 0$).

For the sake of convenience we shall refer to  ``surface density''  as ``density''  
for the remainder of the paper.
%%%%SFRs VS. H\,{\scriptsize I}%%%%%
\subsection{$\Sigma_{\rm H\,{\scriptsize I}}$ vs. ${\rm \Sigma_{SFR}}$}
\begin{figure}[h]
\includegraphics[scale=.50]{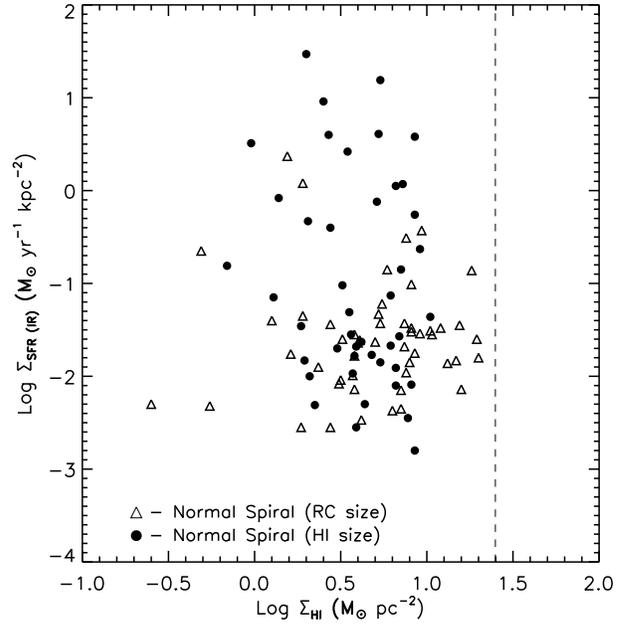}
\caption{${\rm \Sigma_{SFR}}$ vs. $\Sigma_{\rm H\,{\scriptsize I}}$ for the 95 normal disk 
	galaxies with H\,{\scriptsize I} measurements. Filled circles indicate galaxies for which 
		$\Sigma_{\rm H\,{\scriptsize I}}$ was derived by averaging total H\,{\scriptsize I}
		mass over H\,{\scriptsize I} size, while triangles
	show galaxies where $\Sigma_{\rm H\,{\scriptsize I}}$ was determined by averaging 
	H\,{\scriptsize I} mass within the RC size (see \S~\ref{subsubsection:HI-data}).  
	The plot shows no correlation between ${\rm \Sigma_{SFR}}$ and $\Sigma_{\rm H\,{\scriptsize I}}$.
        The dashed line indicates the maximum  H\,{\scriptsize I} surface density of 
		25 ${\rm M_\odot~pc^{-2}}$ found in  our sample.}
\label{figure:HI_IR}
\end{figure}

Fig.\ \ref{figure:HI_IR} shows the distribution of ${\rm \Sigma_{SFR}}$ versus ${\rm \Sigma_{H\,{\scriptsize I}}}$
for the 95 normal disk galaxies with H\,{\scriptsize I} density measurements 
from 21 cm maps (see \S~\ref{subsubsection:HI-data}).
We find there is no correlation between the globally averaged densities of SFRs and
H\,{\scriptsize I} in disk galaxies. The values of ${\rm \Sigma_{H\,{\scriptsize I}}}$ 
stays within a fairly small range ($\le$ 2 dex),
while ${\rm \Sigma_{SFR}}$ spans more than 4 orders of magnitude ($\sim$ 5 dex).

Another feature in Fig.\ \ref{figure:HI_IR} is that the H\,{\scriptsize I}
density saturates at a maximum value of about 25 ${\rm M_\odot~pc^{-2}}$
(indicated by the dashed line).  A similar cutoff in H\,{\scriptsize I} density
was also seen in the studies on spatially resolved SF law
\citep{wong2002,kennicutt2007,bigiel2008}. The saturation in H\,{\scriptsize I}
density may  indicate a transition  from
atomic to molecular gas  where at
increasing column densities, the cold atomic gas  habors regions that  are able to
self-shield against photo-dissociation by the interstellar
radiation field (ISRF) and thus  allow  H$_2$ to
form \citep{krumholz2008,krumholz2009}. 

%%%%SFRs VS. H2%%%%%
\subsection{$\Sigma_{\rm H_2}$ vs. ${\rm \Sigma_{SFR}}$}
\subsubsection{Fixed $\alpha_{\rm CO}$}
Fig.\ \ref{figure:KSL_H2} (a) shows the relation between ${\rm \Sigma_{SFR}}$
and $\Sigma_{\rm H_2}$ on a logarithmic scale for all galaxies in our sample,
where a fixed Galactic CO-to-${\rm H_2}$ conversion factor ($\alpha_{\rm CO}$) value of 4.6 ${\rm
M_\odot~(K~km~s^{-1}~pc^2)^{-1}}$ was adopted \citep{young1991}.  Taken
together, the normal disks and (U)LIRGs span a dynamic range of approximately
$10^6$ in SFR and molecular gas  densities.  There is a strong
correlation between ${\rm \Sigma_{SFR}}$  and $\Sigma_{\rm H_2}$ and a 
Spearman rank correlation test gives a correlation coefficient of 0.96.
Fitting the power law relation using {\tt linmix\_err} 
to the data (with associated uncertainties of 50\%)  yields $N = 1.14 \pm 0.02$
and $A = -3.16 \pm 0.06$ (solid line in Fig.\ \ref{figure:KSL_H2} (a)). The
rms scatter of the data around this best-fit line is 
$\sim 0.40$ dex.  Fitting to disk galaxies and (U)LIRGs separately yields slopes
of $N = 0.97 \pm 0.04$ and $N = 1.12 \pm 0.04$, respectively.  The
scatter of the data around these best fit lines is significantly larger
 for the disk galaxies  ($\sim 0.41$ dex)  than for the (U)LIRGs ($\sim 0.22$ dex).  This
is partly due to the shallower ${\rm \Sigma_{SFR}}$ - $\Sigma_{\rm H_2}$
relation  at low  densities, where the  (formal) ${\rm \Sigma_{SFR}}$-to-$\Sigma_{\rm H_2}$ ratios
 exceed that expected from the best fit. 

\subsubsection{Varying $\alpha_{\rm CO}$}\label{subsubsection:varying-CO-factor}
\begin{figure}[h]
\includegraphics[scale=0.4]{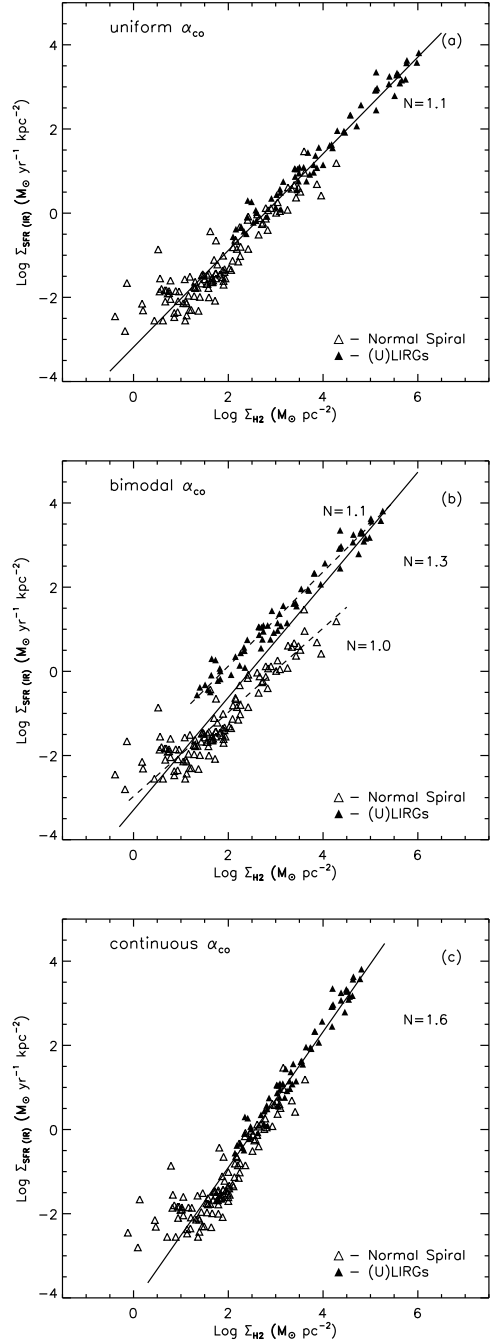}
\caption{${\rm \Sigma_{SFR}}$ vs. $\Sigma_{\rm H_2}$ in the cases where (a) a single 
(Galactic) $\alpha_{\rm CO}$=4.6 value has been applied to the entire sample;
(b) separate values of $\alpha_{\rm CO}$ = 4.6 and 0.8 were applied to disk and (U)LIRGs,
respectively, and (c) a continuously varying $\alpha_{\rm CO}$ was applied.
}
\label{figure:KSL_H2}
\end{figure}

 In the following, we examine the SF law 
obtained when adopting a varying, rather than a fixed,
CO-to-H$_2$ conversion factor.  Our first approach is to adopt 
a conversion factor of 0.8 ${\rm M_\odot~(K~km~s^{-1}~pc^2)^{-1}}$ for
(U)LIRGs \citep{downes1998}, but maintaining the Galactic value for normal disk
galaxies.  As a result, normal galaxies and (U)LIRGs split up into two distinct
${\rm \Sigma_{SFR}}$ vs.\  $\Sigma_{\rm H_2}$ relations as  shown in Fig.\ \ref{figure:KSL_H2} (b).
The fitted  slopes for each sub-sample, i.e., normal galaxies and (U)LIRGs, do not change,
but the  normalization ($A$) for (U)LIRGs increases by $\sim$0.8
dex,  indicating the increased SF efficiencies in these systems.

Our second approach is to adopt continuously varying $\alpha_{\rm CO}$-values
based on the modeling results by \citet{narayanan2012} who parametrizes
$\alpha_{\rm CO}$ as a function of CO line intensity and metallicity (see their
eq.\ 11). Assuming a constant metallicity $Z = 1$ for all  our
galaxies, we found  $\alpha_{\rm CO}$ values in the
range  $3.5 - 6.3~{\rm M_\odot~(K~km~s^{-1}~pc^2)^{-1}}$ for the low-luminosity galaxies,  
$2.0 - 5.0~{\rm M_\odot~(K~km~s^{-1}~pc^2)^{-1}}$ for normal spirals and  $0.3 - 3.0~{\rm
M_\odot~(K~km~s^{-1}~pc^2)^{-1}}$ for (U)LIRGs.  
The resulting $\rm \Sigma_{SFR}$-$\Sigma_{\rm H_2}$ relation
is shown in Fig.\ \ref{figure:KSL_H2} (c) and is seen to be significantly steeper
($N\simeq 1.6$) compared to the relations in Fig.\ \ref{figure:KSL_H2} (a) and (b).
A best-fit power-law to the full sample yields a unimodal ${\rm
\Sigma_{SFR}}$-$\Sigma_{\rm H_2}$ relation of the form: ${\rm \log
\Sigma_{SFR}=(1.60 \pm 0.03) \log \Sigma_{H_2}-(4.11 \pm 0.09)}$.

%%%%SFRs VS. gas%%%%%
\subsection{${\rm \Sigma_{gas}}$ vs. $\Sigma_{\rm SFR}$}
\begin{figure}[h]
	\includegraphics[scale=.40]{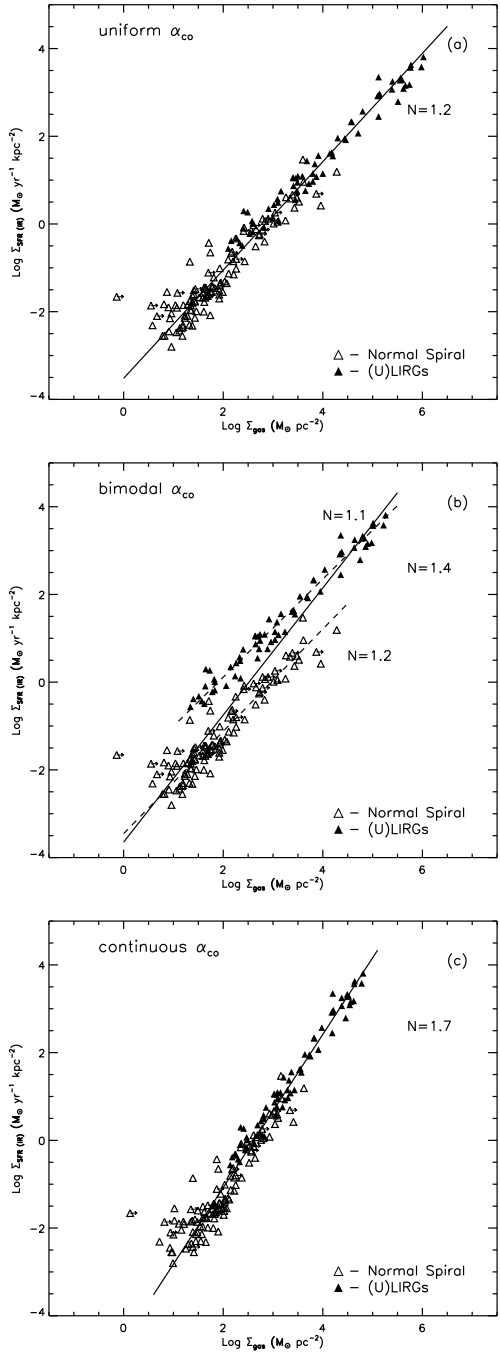} 
\caption{${\rm \Sigma_{SFR}}$ vs. $\Sigma_{\rm gas}$ in the cases where (a) a single 
(Galactic) $\alpha_{\rm CO}$=4.6 value has been applied to the entire sample;
(b) separate values of $\alpha_{\rm CO}$ = 4.6 and 0.8 were applied to disk and (U)LIRGs,
respectively, and (c) a continuously varying $\alpha_{\rm CO}$ was applied.
We took values of $\Sigma_{\rm H_2}$ as lower limits on $\Sigma_{\rm gas}$
in the cases where $\Sigma_{\rm H_I}$ were unavailable.
The $\Sigma_{\rm SFR}$-${\rm \Sigma_{gas}}$ relations generally follow the 
${\rm \Sigma_{SFR}}$-$\Sigma_{\rm H_2}$ relations, but exhibiting steeper slopes.}
\label{figure:KSL_gas}
\end{figure}

The total gas density $\Sigma_{\rm gas}$ is the sum of atomic gas density
$\Sigma_{\rm H\,{\scriptsize I}}$ and molecular gas density $\Sigma_{\rm H_2}$.
When $\Sigma_{\rm H\,{\scriptsize I}}$ data were unavailable  (which was the case for a few normal galaxies), we took
$\Sigma_{\rm H_2}$ as a lower limit on $\Sigma_{\rm gas}$.  For (U)LIRGs, the
atomic gas content is often negligible compared with the molecular gas, and we
set $\Sigma_{\rm gas} \simeq \Sigma_{\rm H_2}$.  Fig.\ \ref{figure:KSL_gas}
(a-c) shows $\Sigma_{\rm SFR}$ vs.\ $\Sigma_{\rm gas}$, where a uniform,
bi-modal and continuous $\alpha_{\rm CO}$  has been adopted, respectively.

We derive the following best-fits to the data:\newline
uniform $\alpha_{\rm CO}$:
\begin{eqnarray}
    \Sigma_{\rm SFR} &=& 10^{-3.56 \pm 0.06}\Sigma_{\rm gas}^{1.24 \pm 0.02}~~~{\rm (disks+(U)LIRGs)}
\end{eqnarray}
bi-modal $\alpha_{\rm CO}$:
\begin{eqnarray}
	\Sigma_{\rm SFR} &=& 10^{-3.46 \pm 0.09}\Sigma_{\rm gas}^{1.15 \pm 0.04}~~~{\rm (disks)}\\
	\Sigma_{\rm SFR} &=& 10^{-2.12 \pm 0.13}\Sigma_{\rm gas}^{1.12 \pm 0.04}~~~{\rm ((U)LIRGs)}\\
    \Sigma_{\rm SFR} &=& 10^{-3.73 \pm 0.11}\Sigma_{\rm gas}^{1.46 \pm 0.04}~~~{\rm (disks+(U)LIRGs)}
\end{eqnarray}
continuous $\alpha_{\rm CO}$:
\begin{eqnarray}
    \Sigma_{\rm SFR} &=& 10^{-4.61 \pm 0.09}\Sigma_{\rm gas}^{1.75 \pm 0.03}~~~{\rm (disks+(U)LIRGs)}
\end{eqnarray}
The scatter of the data around the above best-fit relations are
$\sim 0.36$ dex, $\sim 0.62$ dex (disks+(U)LIRGs), and $\sim 0.43$ dex,
respectively.

The ${\rm \Sigma_{SFR}}$ is a steeper function of total gas density $\Sigma_{\rm
gas}$ than that of molecular gas density $\Sigma_{\rm H_2}$.  This is mainly
owing to the increasing fraction of atomic gas towards lower luminosity
galaxies. At the high density end (mostly populated by molecular-rich spirals
and (U)LIRGs), the ISM is largely molecular and $\Sigma_{\rm SFR}-\Sigma_{\rm
gas}$ essentially follows the $\Sigma_{\rm SFR} - \Sigma_{\rm H_2}$ relation.

%%%%SFRs VS. dense%%%%%
\subsection{${\rm \Sigma_{dense}}$ vs. $\Sigma_{\rm SFR}$} 
\begin{figure}[h]
\includegraphics[scale=.50]{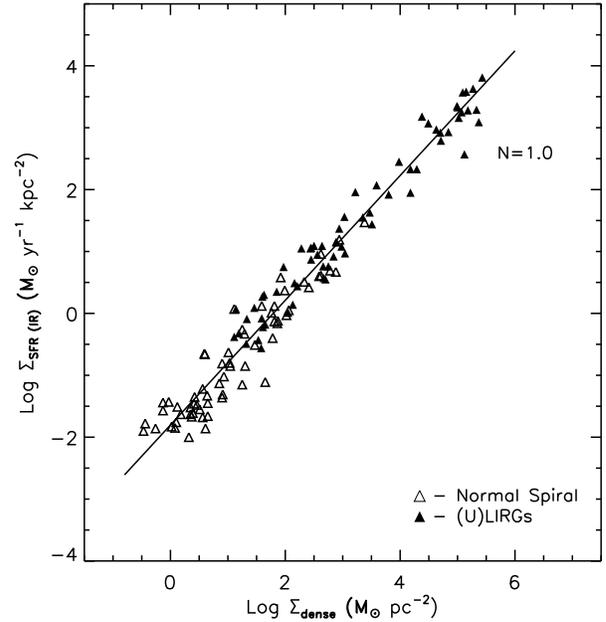}
\caption{${\rm \Sigma_{dense}}$ vs. $\Sigma_{\rm SFR}$ for our sample. The ${\rm \Sigma_{dense}}$ 
- $\Sigma_{\rm SFR}$ relation is the tightest one among all correlations, 
and it is linear (N=$1.01 \pm 0.02$). Fitting to normal galaxies and (U)LIRGs separately 
yields identical linear correlation within the uncertainties.} 
\label{figure:dense_IR}
\end{figure}

Fig.\ \ref{figure:dense_IR} shows $\Sigma_{\rm SFR}$ vs.\ $\Sigma _{\rm dense}$,
where $\Sigma _{\rm dense}$ was derived for our compiled HCN sample of 128
galaxies adopting a constant Galactic ${\rm HCN-to-M_{dense}}$ conversion factor
($\alpha^{\rm MW}_{\rm HCN}$) value of 10 ${\rm M_\odot~(K~km~s^{-1}~pc^2)^{-1}}$.  One
can clearly see a very tight $\Sigma_{\rm SFR}$-${\rm \Sigma_{dense}}$
correlation spanning more than $\sim 6$ orders of magnitude in density
 and we find a best-fit log-linear relation of:  
\begin{equation}
{\rm \log \Sigma_{SFR} = (1.01 \pm 0.02) \log{\Sigma_{dense}} - (1.82 \pm 0.05)}.
\end{equation}
A Spearman rank correlation test  yields a correlation coefficient of $\sim 0.99$.  The
dispersion in the $\Sigma_{\rm SFR}$-${\rm \Sigma_{dense}}$ relation ($\sim
0.32$ dex) is  smaller than the relations between densities
of SFRs and other gas components (atomic, molecular, and total gas).

We find a linear SF law ($N=1.01 \pm 0.02$) in terms of dense molecular gas, in
agreement with the tight linear relation found between ${\rm L_{IR}}$ and ${\rm L_{HCN}}$ (GS04a,b).
Scaling the dense molecular gas mass and SFRs by our measured RC
disk sizes does not change the slope of the ${\rm L_{IR}}$-${\rm L_{HCN}}$ correlation since this
relation is essentially exactly linear. 

Fitting to spiral galaxies and (U)LIRGs separately (each constituting about half of the HCN
sample), we derive identical ${\rm \Sigma_{SFR}}-\Sigma_{\rm dense}$ relations within the errors, 
both in terms of the slope ($N$) and the overall normalization ($A$): $N=0.98 \pm 0.05$ ($N=0.96
\pm 0.04$) and $A=-1.84 \pm 0.06$ ($A=-1.57 \pm 0.10$) for normal spirals ((U)LIRGs).

So far, we have only presented the results of the various SF laws related to IR-based SFRs.  As
mentioned above, we derived our SFRs in two independent ways using both IR and RC data.  Excellent
agreement were found between the two SFRs estimators, and using one over the other, therefore, did
not alter our results in any significant way. Further details on the various SF laws related to
RC-based SFRs are given in the Appendix \ref{appendix:SFRs}. 

%%%%%%%%%%%%%%%%%%
%  Discussion and Analysis 
%%%%%%%%%%%%%%%%%%
\section{Discussion} \label{section:discussion}

%%%%%%%%%%%%%%%%H2%%%%%%%%%%%%%%%%%%%%%%
\subsection{Deviations from the molecular gas SF law}
An interesting finding in our global ${\rm \Sigma_{SFR}}$ - $\Sigma_{\rm H_2}$
relation is the higher IR-to-CO ratios  exhibited by less massive, low-luminosity
galaxies compared with normal spirals (Fig.\ \ref{figure:KSL_H2}). 
These galaxies have $\Sigma_{\rm H_2} \le 10~{\rm M_\odot~yr^{-1}}$
and deviate significantly from the derived ${\rm \Sigma_{SFR}}$ - $\Sigma_{\rm H_2}$
relations for whole sample (see solid lines in Fig.\ \ref{figure:KSL_H2}).
These low-luminosity galaxies are usually atomic gas dominated and metal-poor 
\citep{sadler2000,sadler2001}.
The higher IR-to-CO ratios in these galaxies may be 
due to systematic underestimates of their total molecular gas mass as 
the adopted $\alpha_{\rm CO}$ are too small for these systems.
In low-metallicity environments, photodissociation destroys CO more easily, 
whereas the H$_2$ can self-shield more effectively \citep{narayanan2012}
so that the $\alpha_{\rm CO}$ tends to be larger than the Galactic value 
and increases with decreasing metallicity and dust-to-gas ratio
\citep[e.g.,][]{boselli2002,leroy2011,krumholz2011,rahman2012,bolatto2011,bolatto2013}.  
Although a continuously varying $\alpha_{\rm CO}$ was introduced in 
our analysis, as we assumed constant metallicity $Z=1$ (\S~\ref{subsubsection:varying-CO-factor})
we could still underestimate their molecular gas mass.

Alternatively, we note that, some low-luminosity galaxies may
indeed exhibit higher SFE than large spiral galaxies \citep{gardan2007,gratier2010,schruba2011,
saintonge2012,leroy2013}. Minor starburst events in these systems \citep{saintonge2012}, or
suppression of SF in large spiral galaxies by dynamical effects \citep{leroy2013}, 
are possible explanations.

At the high density end of Fig.\ \ref{figure:KSL_H2}, 
where mostly (U)LIRGs (mergers) reside, the molecular SFE goes up again as the plot
turns steeper in these regimes. 
In these galaxy mergers, the observed $\alpha_{\rm CO}$ are suggested to
be much lower (by a typical factor of $\sim~2-10$) than the normal galaxies because the 
rise in velocity dispersion and kinetic temperature in GMCs 
during the merger increases the CO intensity \citep{narayanan2011}.
The ${\rm \Sigma_{SFR}}$ - $\Sigma_{\rm H_2}$ relation derived by using bimodal 
$\alpha_{\rm CO}$ clearly display the higher ${\rm \Sigma_{SFR}}$-to-$\Sigma_{\rm H_2}$ 
ratios of (U)LIRGs than that of normal spirals, showing two distinct regimes, 
one for normal galaxies and one for starbursts (see Fig.\ \ref{figure:KSL_H2} (b)).
It appears to be similar to the bi-modal relations found from high-$z$ CO observations,
where two different $\alpha_{\rm CO}$ conversion factors were used
\citep{daddi2010,genzel2010,ivison2011,riechers2012}.

The trend that (U)LIRGs have higher molecular SFE is even more obvious as
a continuously varying $\alpha_{\rm CO}$ was adopted. 
We found a unimodal ${\rm \Sigma_{SFR}}$-$\Sigma_{\rm H_2}$ relation with 
a clearly super-linear power-law index ($N=1.60\pm0.03$, see Fig.\ \ref{figure:KSL_H2} (c)).
It is worth noting here that a metallicity dependence would further steepen this fit.
We noticed that this power law relation is comparable to that
derived by \citet{bouche2007} which obtained a universal SF
law for a sample of local and high-$z$ galaxies 
(${\rm \log \Sigma_{SFR}=(1.7 \pm 0.05) \log \Sigma_{H_2}+(-4.03 \pm 0.10)}$).
We tend to believe that $\alpha_{\rm CO}$ varies continuously
rather than in a bi-model way because there should be no clear 
boundary between starbursts and disk spirals.
Our RC images show that some normal galaxies exhibit strong compact starburst 
regions in the center which are comparable to 
many LIRGs (such as NGC253, NGC1097 and NGC3504).
As a consequence, the derived densities for disk galaxies and (U)LIRGs
show an overlapping region up to two orders of magnitude (see Fig.\ \ref{figure:KSL_H2}
\& Fig.\ \ref{figure:KSL_gas}).
It is thus natural to expect the $\alpha_{\rm CO}$ varies continuously
from normal disks to (U)LIRGs as well.
A super-linear power law index of $\sim 1.6$ derived here suggested a density-dependent SFE.

Although the result is sensitive to the $\alpha_{\rm CO}$ conversion factor,
we still get a general picture of SF law in molecular gas:
in disk galaxies with $10 < \Sigma_{\rm H_2} < 100 {\rm~M_\odot~pc^{-2}}$, 
where molecular gas dominates and is gravitationally
bound to GMCs with approximately uniform properties, 
the ${\rm \Sigma_{SFR}}$ is roughly linear proportional to $\Sigma_{\rm H_2}$.
The value of $\Sigma_{\rm H_2} \sim 100 {\rm~M_\odot~pc^{-2}}$ is 
the characteristic density of individual giant molecular clouds (GMCs).
In starburst galaxies (mostly (U)LIRGs) with $\Sigma_{\rm H_2} > 100 {\rm~M_\odot~pc^{-2}}$,
where supersonic turbulence is known to be a dominant process controlling SF, 
the ${\rm \Sigma_{SFR}}$-$\Sigma_{\rm H_2}$ relation appears considerably 
steeper than linear, implying an increasing molecular SFE.
In low-luminosity galaxies with $\Sigma_{\rm H_2} < 10 {\rm~M_\odot~pc^{-2}}$, 
where atomic gas dominates ISM, 
the ${\rm \Sigma_{SFR}}$-$\Sigma_{\rm H_2}$ relation show visibly increasing
galaxy-to-galaxy scatter and higher IR-to-CO ratios.

\subsection{The role of atomic gas and SF law for total gas}
%%%%%%%%%%%%%%%%%%%%%%%%%%%%%H\,{\scriptsize I}%%%%%%%%%%%%%
We found there is no correlation between ${\rm \Sigma_{SFR}}$ and 
$\Sigma_{\rm H\,{\scriptsize I}}$ (see Fig.\ \ref{figure:HI_IR}).
We can even see the discrepancy between the morphologies of SFR and atomic gas 
from RC maps and H\,{\scriptsize I} maps, that some galaxies
show central H\,{\scriptsize I} depressions where significant SF is still occurring.
More detailed studies on local scales showed little or no correlation between ${\rm \Sigma_{SFR}}$ 
and $\Sigma_{\rm H\,{\scriptsize I}}$ as well \citep{wong2002,kennicutt2007,bigiel2008}.

The H\,{\scriptsize I} saturation value of $\sim 25~{\rm M_\odot~pc^{-2}}$ found in our study 
coincides with the maximum H\,{\scriptsize I} value derived from \citet{kennicutt2007} studying 
the spatially resolved SF law in M51 on $\sim$500 pc scale,
but is higher than the value of ${\rm \sim 10~M_\odot~pc^{-2}}$ found by 
\citet{wong2002} and \citet{bigiel2008}.
We supposed this may due to the larger range of metallicity in our sample,
given the result by theoretical modeling 
that the characteristic column density of the transition from 
atomic to molecular gas increases with metallicity, 
and is around ${\rm \sim 10~M_\odot~pc^{-2}}$ at solar metallicity \citep{krumholz2009}. 

%%%%%%%%%%%%%%%%%%%%%%%%Gas%%%%%%%%%%%%%%%%%%%%%
Although atomic gas itself shows no correlation with SFR, but it still
plays important role in regulating the SF law of total gas.
The combined density of atomic and molecular gas, i.e., the total gas density
${\rm \Sigma_{gas}}$, shows a steeper function with ${\rm \Sigma_{SFR}}$
than molecular gas density ${\rm \Sigma_{H_2}}$.
This is mainly due to the increasing fraction of atomic gas towards 
lower luminosity galaxies.
An interesting question is whether total gas density ${\rm \Sigma_{gas}}$
or molecular gas density ${\rm \Sigma_{H_2}}$ correlates 
better with the star formation ${\rm \Sigma_{SFR}}$.
That is which of the ${\rm \Sigma_{SFR}}$ - $\Sigma_{\rm H_2}$ 
and ${\rm \Sigma_{SFR}}$ - $\Sigma_{\rm gas}$ relations might be more fundamental. 
Some works asserted that the good correlation between ${\rm \Sigma_{SFR}}$ 
and $\Sigma_{\rm gas}$ is driven by the molecular gas SF law
and the ratio of molecular to atomic hydrogen
\citep{wong2002,bigiel2008,blanc2009}.
In this picture, molecular clouds are first formed from H\,{\scriptsize I},
and stars are then formed exclusively from molecular gas.
On the contrary, some other works claimed that it is
the gas phases as a whole feeds the SF activity.
Where the gas is predominantly molecular in most of spiral galaxies and (U)LIRGs, 
the correlation of ${\rm \Sigma_{SFR}}$ with  ${\rm \Sigma_{H_2}}$ could be only a 
manifestation of the underlying correlation of the ${\rm \Sigma_{SFR}}$ 
with  ${\rm \Sigma_{gas}}$ \citep{kennicutt2007,fumagalli2009}.
Our work shows that at the density regimes with ${\rm \Sigma_{H_2} > 10~M_\odot~pc^{-2}}$,
the ${\rm \Sigma_{SFR}}$ correlates equally well with total gas 
${\rm \Sigma_{gas}}$ and molecular gas ${\rm \Sigma_{H_2}}$.
While at the low density regime where ${\rm \Sigma_{H_2} < 10~M_\odot~pc^{-2}}$, 
we can see that the scatter in the $\Sigma_{\rm SFR}$ - $\Sigma_{\rm gas}$ relation
is significantly smaller (with scatter $\sim 0.43$ dex, see Fig.\ \ref{figure:KSL_H2}) 
than that in the ${\rm \Sigma_{SFR}}$ - $\Sigma_{\rm H_2}$ relation (with scatter
$\sim 0.55$ dex, see Fig.\ \ref{figure:KSL_gas}). 
This may indicate that there could be some other parameter apart from 
the molecular gas content that determines the SFRs for galaxies, 
at least in low-luminosity galaxies, and atomic gas may have indirect effect on SF.

Several  investigations of the total gas SF law have found a 
 threshold  in the gas surface density  (${\rm \Sigma_{gas} \sim 5-10~M_\odot~pc^{-2}}$)
below which  the  SFR declines  rapidly and SF is strongly suppressed
\citep{kennicutt1998,martin2001,leroy2005,kennicutt2008}.  These thresholds are
mostly observed in galaxies like dwarf galaxies and low surface brightness (LSB)
galaxies \citep{leroy2005,wyder2009}.  As most of our galaxies have total gas
density larger than the characteristic threshold value of ${\rm \sim
10~M_\odot~pc^{-2}}$ (with less than 10 galaxies having ${\rm \Sigma_{SFR}}$
under this value), we do not find a clear threshold in our sample.

\subsection{SFRs vs dense gas}
%%%%%%%%%%%%%%%%%%%%%%%%%%%Dense Gas%%%%%%%%%%%%%%%%%%%%%%%%%%%
The tightest relation (with scatter of $\sim 0.32$ dex) in our results is the linear relation 
between ${\rm \Sigma_{SFR}}$ and ${\rm \Sigma_{dense}}$.
We fit the normal galaxies and (U)LIRGs separately and find 
both the power indices are close to unity.
Therefore, the tight linear correlation between ${\rm \Sigma _{dense}}$
and ${\rm \Sigma_{SFR}}$ is established within the whole sample of galaxies, 
independent of IR luminosity.

In contrast, \citet{GB-2012} suggested a two-function power law 
in terms of dense molecular gas SF law,
which claimed a higher slope in (U)LIRGs ($1.05\pm0.05 $) 
than normal galaxies ($0.90\pm0.06 $).
We noticed that \citet{GB-2012} adopted a bi-modal
HCN-to-${\rm M_{dense}}$ conversion factor $\alpha_{\rm HCN}$,
that $\alpha_{\rm HCN}$ is $\sim 3$ times lower in (U)LIRGs.
To test the effect of this varying $\alpha_{\rm HCN}$ on the dense gas SF law, 
we applied a value of  $1/2\alpha^{\rm MW}_{\rm HCN}$ 
to LIRGs ($10^{11} {\rm L_\odot} \le L_{\rm FIR}
\le 10^{12} {\rm L_\odot}$) and a value of 
$1/3\alpha^{\rm MW}_{\rm HCN}$ to (U)LIRGs ($L_{\rm FIR} \ge 10^{12} {\rm L_\odot}$).
In this  case, we found the power-index of $\Sigma_{\rm SFR}$-${\rm \Sigma_{dense}}$
relation is only slightly changed from $1.01 \pm 0.02$ to $1.09 \pm 0.02$.
The slopes fitted for normal galaxies and (U)LIRGs separately are 
$0.98 \pm 0.05$ and $0.97 \pm 0.03$, respectively,
agreeing well with each other within errors. 

We  notice that \citet{GB-2012} utilized  sizes of the star forming  regions
determined from optical/NIR images of hydrogen recombination lines (${\rm
H_{\alpha}}$ and ${\rm P_{a\alpha}}$), which  suffer
from heavy dust extinction especially in the dense H\,{\scriptsize II} regions
in starbursts.  We  find that the ${\rm
H_{\alpha}}$/${\rm P_{a\alpha}}$ sizes of \citet{GB-2012}  are
roughly $2-4\times$ smaller than our RC  sizes  for the same luminous 
infrared galaxies.  We suspect that
this could be the main reason to cause the discrepancy between their results
with ours.  Besides, the (U)LIRGs sample of \citet{GB-2012} includes also
high-$z$ galaxies (mostly QSOs), whose IR luminosities are likely to be
contaminated by AGNs.  This could be another reason for the steeper slope of
(U)LIRGs found in their work.  Another factor that might affect ${\rm
\Sigma_{SFR}}$/$\Sigma_{\rm dense}$ is the FIR 60-to-100 $\mu m$ color [C(60/100)].
GS04b found a weak correlation between $L_{\rm IR}$/$L_{\rm HCN}$ and
the FIR color, which indicated that the warm dust temperature $T_{\rm dust}$
also plays a role in the ratio of SFRs-to-dense gas.
More details related to FIR 60-to-100$\mu m$ color are given in
the Appendix \ref{appendix:FIR color}.

%%%%%%%%%%%%%%%%%%%%%%%%%%%%%%%%%%%%%%%%%%%%%%
%       Toward a better understanding of SF law
%%%%%%%%%%%%%%%%%%%%%%%%%%%%%%%%%%%%%%%%%%%%%%
\subsection{Toward a better understanding of SF law}

%%%%%%%%%%1. Comparison with Spatially Resolved Results
Recent studies of  spatially resolved SF  laws in nearby star forming galaxies at  
scales of $\sim 1$ kpc or less found linear ${\rm
\Sigma_{SFR}}$-$\Sigma_{\rm H_2}$ relations  with typical scatter in the range $0.2 - 0.3$
dex \citep{kennicutt2007,blanc2009,rahman2011,verley2010,
bigiel2008,bigiel2011, schruba2011,rahman2012}.  This  includes  azimuthally-averaged
studies \citep{wong2002,murgia2002,schuster2007} and pixel-by-pixel
 analyses \citep{kennicutt2007,bigiel2008,bigiel2011,leroy2008,leroy2013,schruba2012}.

Although this study examines globally-averaged properties , 
our sample spans a  larger range ($\sim 10^6$) in densities of both
 SFR and molecular gas than previous resolved studies
($\sim 10^{2-4}$).  We found an approximately linear ${\rm
\Sigma_{SFR}}$-$\Sigma_{\rm H_2}$ relation as well (using constant $\alpha_{\rm
CO}$).  However, we found much larger scatter in our derived global ${\rm
\Sigma_{SFR}}$-$\Sigma_{\rm H_2}$ relations (with scatter $\sim 0.40$
dex), especially toward low luminosity end where atomic gas H\,{\scriptsize I},
rather than ${\rm H_{2}}$, is dominated in the disks of galaxies (with
scatter $\sim 0.55$ dex, see Fig.\ \ref{figure:KSL_H2} \&
\ref{figure:KSL_gas}).  This discrepancy implies that the scatter in the
observed relations between ${\rm \Sigma_{SFR}}$ and $\Sigma_{\rm H_2}$ is mainly
driven by galaxy-to-galaxy variations, particularly reflected in low-luminosity
galaxies which have low-metallicity and are dominated by atomic gas.  Indeed,
\citet{schruba2011}, which studied the spatially resolved  ${\rm
\Sigma_{SFR}}$-$\Sigma_{\rm H_2}$ relation in the atomic gas dominated regime in
nearby galaxies using stacking technique, found a much larger scatter in ${\rm
\Sigma_{SFR}}$-$\Sigma_{\rm H_2}$ relation compared with previous studies on
molecular gas rich regions.  But once the galaxy-to-galaxy scatter was removed,
they found a remarkably tight, linear ${\rm \Sigma_{SFR}}$-$\Sigma_{\rm H_2}$
relation (with scatter around $\sim 0.25$ dex).

It is interesting to discuss why the global ${\rm \Sigma_{SFR}}$-$\Sigma_{\rm H_2}$
relation exhibits much larger scatter than the resolved  relations.
The dispersion of $\alpha_{\rm CO}$ values among different type of
galaxies seems to be usually larger than that within a single galaxy.
A generally flat profile of $\alpha_{\rm CO}$ as a function of 
galactocentric radius was observed (except that the central $\alpha_{\rm CO}$
value with $\sim 1$ kpc can be a few of factors below)  \citep{sandstrom2013,narayanan2012}.
Secondly, some part of the global scatter could be physically real, 
possibly reflecting the unknown plane thickness variation, different stellar mass,
dust abundances, metallicity and SF histories from galaxy to galaxy,
or any further parameters that are important in setting the global SFRs.
Another important reason could be that the global molecular gas measurements contain
diffuse molecular gas - meaning unbound material and the low density outskirts of bound 
clouds - that extend over whole disks ($\sim {\rm kpc}$) which however could be easily resolved out
by the studies at sub-kpc scale as they often adopted high-resolution 
interferometry measurements .
Since this global diffuse molecular gas is not directly associated with SF activity,
we thus see a weaker ${\rm \Sigma_{SFR}}$-$\Sigma_{\rm H_2}$ correlation in our global studies.

Several  studies in literature
have derived a variety of power-law indices for the ${\rm
\Sigma_{SFR}}$-$\Sigma_{\rm H_2}$ relation,  ranging from $N<1$
\citep[e.g.,][]{blanc2009},  $N\sim1$
\citep[e.g.,][]{wong2002,bigiel2008},  $N\sim1.3-1.5$
\citep[K98;][]{kennicutt2007,schruba2010} to 
$N\sim1.7-2.0$ \citep{bouche2007,verley2010}. 
A contributing factor to the wide range of observed power-law indices 
stems from the different scales over which surface densities have been
inferred (i.e., global scales, radial profiles, and pixel-by-pixel).
 Within the same individual galaxy, the resolved ${\rm
\Sigma_{SFR}}$-$\Sigma_{\rm H_2}$ relation may depend on the averaging
aperture size on small scales ($\le 300$ pc)
\citep{schruba2010,verley2010,calzetti2012}.  Another contributing factor
is the differences in estimators of SFR and SF size adopted across various
studies
 since  they directly affect the
densities and thus the derived SF laws.  Finally, our work clearly shows
that the ${\rm \Sigma_{SFR}}$-$\Sigma_{\rm H_2}$ relation is sensitive to the
application of $\alpha_{\rm CO}$.

Furthermore, the choice and weighting of sample data also plays a significant role 
in the resulting power indices of SF laws.
 Previous studies on sub-kpc scale demonstrated that the 
${\rm \Sigma_{SFR}}$ - $\Sigma_{\rm H_2}$ relation
is roughly linear in the regime $10 < \Sigma_{\rm H_2} < 100 {\rm~M_\odot~pc^{-2}}$
where normal spiral galaxies are found, and gets super-linear in the 
regime $\Sigma_{\rm H_2} > 100 {\rm~M_\odot~pc^{-2}}$ of starburst galaxies
\citep[e.g.,][]{bigiel2008,leroy2013}.
Our studies on global scales found a similar trend using, for the first time,
RC observations.
In fact, we find that the slopes of ${\rm \Sigma_{SFR}}$ - $\Sigma_{\rm H_2}$ relation
change from $\sim1.0$ to $\sim1.2$ when more and more (U)LIRGs are included 
(when a uniform $\alpha_{\rm CO}$ was used). 
That is the larger weight to the (U)LIRGs sample or the data with 
$\Sigma_{\rm H_2} > 100 {\rm~M_\odot~pc^{-2}}$, 
the steeper slopes in SF laws related to molecular gas and total gas.
This may imply a density-dependent SF efficiency of gas, i.e., 
the higher density exhibits higher SF efficiency. 
A similar trend was also pointed out by GS04b, which studied the global 
IR-CO luminosity correlation of local (U)LIRGs. 
The studies of high-$z$ galaxies indicate a similar behavior as 
well \citep{bouche2007,daddi2010,bothwell2010}.

Our work showed that the linear
$\Sigma_{\rm SFR}$-${\rm \Sigma_{dense}}$ relation is independent of the galaxy
sample and IR luminosity, so that the dense SF law can be described by a
universal one-function law across all galaxy types. 
Not only at global  scales does the SFR density exhibit a tight and
linear correlation with the dense gas density as showed in our work, the HCN
surveys of Galactic dense cores and dense clumps even extend the linear
luminosity correlation down to GMC  scales \citep{wu2005,wu2010}.  
As we adopted lower $\alpha_{\rm HCN}$ for (U)LIRGs ($1/3-1/2\alpha^{\rm HCN}_{\rm MW}$), 
the slope is only slightly changed and still close to unity.  The luminosities of
other dense gas tracer such as HNC(1-0), ${\rm HCO^+}$, CN(2-1) and CS(3-2) were
observed to vary linearly with $L_{\rm FIR}$ as well \citep{baan2008,zhang2014}.

\section{Summary} \label{section:summary}
We have examined the relations between the galaxy-averaged surface
densities of the various gas components (atomic, molecular, total gas and dense
molecular gas) and  SFR in a sample of 181 local galaxies with
IR luminosities spanning $\sim 5$ orders of magnitude (${\rm 10^{7.8}
-10^{12.3}~L_\odot}$). 
We have taken a novel approach  and used high-resolution
radio continuum observations to  accurately measure the
sizes of the areas of active star formation  within the galaxies -- a key step as it directly  
affects the inferred  gas and  SFR surface densities.

We found there is no correlation between ${\rm \Sigma_{SFR}}$ and $\Sigma_{\rm
H\,{\scriptsize I}}$,  and that $\Sigma_{\rm H\,{\scriptsize
I}}$ does not exceed  a maximum value of $\sim 25~{\rm
M_\odot~pc^{-2}}$.  We have explored the  ${\rm
\Sigma_{SFR}}$-$\Sigma_{\rm H_2}$ relation  for different values of the $\alpha_{\rm CO}$ factor.  
Adopting a  Galactic  $\alpha_{\rm CO}$  to all galaxies, a roughly linear power-law index ($N = 1.14
\pm 0.02 $) was found.   If a more realistic $\alpha_{\rm CO}$-value [$0.8\,{\rm M_\odot~(K~km~s^{-1}~pc^2)^{-1}}$]
is applied to the (U)LIRGs in our sample the result is a bimodal SF law, i.e.,
two nearly linear ${\rm \Sigma_{SFR}}$-$\Sigma_{\rm H_2}$ correlations,
indicating that separate SF modes might exist for (U)LIRGs and disk galaxies.
A unimodal, but significantly steeper ($N = 1.62 \pm 0.04 $), 
${\rm \Sigma_{SFR}}$-$\Sigma_{\rm H_2}$ relation is recovered if a
continuously-varying $\alpha_{\rm CO}$ conversion factor is applied.
  The global
$\Sigma_{\rm SFR}$ shows steeper functions of total gas density $\Sigma_{\rm
gas}$ than those of $\Sigma_{\rm H_2}$ (the slopes depend on the $\alpha_{\rm
CO}$ as well), and are tighter than $\Sigma_{\rm SFR}$-$\Sigma_{\rm H_2}$
relation among low-luminosity galaxies.  The dense molecular gas density
$\Sigma_{\rm dense}$ shows the best correlation with $\Sigma_{\rm SFR}$, which
is exactly linear with index $N=1.01 \pm 0.02$.  Fitting the normal galaxies and
(U)LIRGs separately yields the same linear slopes.

Out of these results the following picture emerges. The SF is more directly
related to molecular gas than atomic gas.  But the seeming roughly linear
relation of ${\rm \Sigma_{SFR}}$-$\Sigma_{\rm H_2}$ can easily gets super-linear
as a varying-$\alpha_{\rm CO}$ was adopted and the number of (U)LIRGs increases.
The  fact that the low-luminosity galaxies  lie significantly above the 
${\rm \Sigma_{SFR}}$-$\Sigma_{\rm H_2}$ relation derived for whole sample
suggest there could be some other parameter apart from
the molecular gas content that determines the SF in these galaxies.
Unlike the molecular SF law, the linear $\Sigma_{\rm SFR}$-${\rm
\Sigma_{dense}}$ relation is independent of the choice of sample (thus galaxy
luminosity)  and the adopted $\alpha_{\rm HCN}$ factor, suggesting the basic
units of SF in galaxies are in dense cores.

\bigskip
This work is partially supported by NSFC grant Nos. 11173059 and 11390373,
and CAS No. XDB09000000. This research has made use
of the NASA IPAC Extragalactic Database (NED) which is
operated by the Jet Propulsion Laboratory, California Institute
of Technology, under contract with the National Aeronautics
and Space Administration.

\clearpage

%%%%%%%%%%%%%%%%%%%%%%%%%%%%%%%%%%%%%%%
%                 Table 1
%%%%%%%%%%%%%%%%%%%%%%%%%%%%%%%%%%%%%%%
\LongTables 
\begin{deluxetable*}{lllllll||lllllll} 
\tabletypesize{\tiny}
\renewcommand{\arraystretch}{0.4}
\setlength{\tabcolsep}{1pt}
\tablecolumns{14}
\tablewidth{0pc}
\tablecaption{Luminosities \label{table:table1}}
\tablehead{ \colhead{Name} & \colhead{D} & \colhead{$L_{\rm CO}$} & \colhead{$L_{\rm HCN}$} 
 & \colhead{$L_{\rm IR}$} & \colhead{$L_{\rm RC}$} & \colhead{Ref.}  
 & \colhead{Name} & \colhead{D}  & \colhead{$L_{\rm CO}$} & \colhead{$L_{\rm HCN}$} 
 & \colhead{$L_{\rm IR}$} & \colhead{$L_{\rm RC}$} & \colhead{Ref.}  \\ }

% & \colhead{\tiny (Mpc)} & \multicolumn{2}{l}{\tiny (${\rm 10^8K~km~s^{-1}~pc^2 }$)} 
% & \colhead{\tiny ($\rm L_{\odot}$)} & \colhead{\tiny (${\rm 10^{21}W~Hz^{-1}}$)} &   
% & \colhead{\tiny (Mpc)} & \multicolumn{2}{l}{\tiny (${\rm 10^8K~km~s^{-1}~pc^2 }$)} 
% & \colhead{\tiny ($\rm L_{\odot}$)} & \colhead{\tiny (${\rm 10^{21}W~Hz^{-1}}$)} &   

\startdata
 &  & \multicolumn{10}{c}{\bf Normal Galaxies} & &  \\
\cline{1-14}  \\
NGC 0134 & 21.38 & 3.76 & 0.4 & 10.62 & 3.48 & 13, 21, 36, 44 &
NGC 0174 & 49.74 & 12.81 & 1.8 & 10.93 & 3.58 & 13, 21, 36, 50 \\
NGC 0253 & 4.11 & 15.26 & 0.66 & 10.69 & 5.13 & 2, 20, 36, 43 &
NGC 0520 & 30.59 & 28.84 & 0.62 & 10.92 & 13.31 & 3, 29, 36, 42 \\
NGC 0598 & 0.81 & 0.06 & \dots & 9.04 & -6.14 & 3, 36, 42 &
NGC 0628 & 10.55 & 6.99 & \dots & 9.99 & -4 & 6, 36, 42 \\
NGC 0660 & 12 & 10 & 0.17 & 10.47 & 0.35 & 3, 20, 36, 42 &
NGC 0772 & 34.52 & 46.93 & \dots & 10.61 & 3.85 & 3, 36, 42 \\
NGC 0891 & 9.17 & 7.68 & 0.22 & 10.33 & 0.65 & 1, 20, 36, 48 &
NGC 0925 & 9.1 & 1.94 & \dots & 9.39 & -5.94 & 3, 36, 48 \\
NGC 0931 & 67.26 & 0.66 & 1.85 & 10.85 & 0.8 & 14, 24, 40, 44 &
NGC 1022 & 19.48 & 3.13 & 0.15 & 10.35 & -4.18 & 1, 20, 36, 44 \\
NGC 1055 & 13.63 & 10.09 & 0.24 & 10.25 & -1.83 & 1, 20, 36, 42 &
NGC 1058 & 8.8 & 0.3 & \dots & 8.93 & -6.31 & 8, 37, 42 \\
NGC 1097 & 16.87 & 4.68 & \dots & 10.71 & 7.73 & 9, 36, 46 &
NGC 1266 & 29.7 & 13.64 & 0.52 & 10.48 & 5.53 & 16, 34, 36, 52 \\
NGC 1530 & 37.01 & 16.08 & 0.53 & 10.71 & 4.63 & 3, 29, 36, 42 &
NGC 1569 & 4.11 & 0.02 & \dots & 9.39 & -5.67 & 3, 36, 42 \\
NGC 1667 & 61.83 & 16.64 & 6.83 & 10.97 & 28.27 & 14, 24, 36, 44 &
NGC 1808 & 12.56 & 6.63 & 0.7 & 10.71 & 3.4 & 11, 21, 36, 46 \\
NGC 2273 & 28.54 & 0.76 & 0.21 & 10.25 & -1.04 & 14, 24, 36, 49 &
NGC 2276 & 36.96 & 9.78 & 0.38 & 10.83 & 41.47 & 1, 20, 36, 42 \\
NGC 2403 & 4.11 & 0.22 & \dots & 9.4 & -5.96 & 3, 36, 42 &
NGC 2764 & 41.36 & 6.17 & 0.1 & 10.33 & -3.76 & 16, 34, 41, 52 \\
NGC 2841 & 13.42 & 8.24 & \dots & 9.72 & -4.62 & 3, 38, 42 &
NGC 2903 & 8.53 & 5.16 & 0.15 & 10.22 & -2.54 & 6, 20, 36, 42 \\
NGC 2976 & 4.11 & 0.25 & \dots & 8.92 & -6.3 & 3, 36, 42 &
NGC 3031 & 4.11 & 0.47 & \dots & 9.57 & -5.63 & 9, 36, 42 \\
NGC 3032 & 25.22 & 1.45 & 0.04 & 9.67 & -5.9 & 15, 32, 41, 52 &
NGC 3079 & 19.51 & 29.68 & 1.22 & 10.79 & 30.98 & 1, 20, 36, 43 \\
NGC 3147 & 43.05 & 61.65 & 0.64 & 10.93 & 15.78 & 3, 29, 36, 42 &
NGC 3310 & 17.06 & 1 & \dots & 10.47 & 0.74 & 3, 36, 42 \\
NGC 3338 & 19.12 & 1.7 & \dots & 9.67 & -5.12 & 3, 37, 42 &
NGC 3351 & 8.71 & 2.8 & \dots & 9.71 & -6.01 & 2, 36, 42 \\
NGC 3368 & 10.42 & 1.62 & \dots & 9.54 & -5.98 & 3, 36, 42 &
NGC 3486 & 8.41 & 0.83 & \dots & 9.22 & -5.94 & 3, 36, 42 \\
NGC 3504 & 26.36 & 10.89 & 0.64 & 10.69 & 16.38 & 2, 25, 36, 42 &
NGC 3521 & 8.28 & 13.27 & \dots & 10.12 & -3.47 & 2, 36, 42 \\
NGC 3556 & 12.09 & 3.69 & 0.1 & 10.24 & -1.44 & 1, 20, 36, 42 &
NGC 3620 & 19.18 & 5.15 & 0.66 & 10.67 & 0.16 & 13, 21, 36, 51 \\
NGC 3627 & 7.34 & 5.9 & 0.05 & 10.1 & -3.48 & 1, 20, 36, 42 &
NGC 3628 & 8.67 & 8 & 0.28 & 10.12 & -2.17 & 1, 20, 36, 43 \\
NGC 3631 & 20.22 & 10.99 & \dots & 10.23 & -2.29 & 2, 36, 42 &
NGC 3665 & 34.83 & 2.56 & 0.12 & 9.97 & 4.34 & 16, 34, 41, 52 \\
NGC 3675 & 11.72 & 4.84 & \dots & 9.85 & -5.64 & 3, 36, 42 &
NGC 3726 & 14.02 & 3.46 & \dots & 9.75 & -5.51 & 3, 36, 42 \\
NGC 3893 & 16.38 & 4.11 & 0.29 & 10.2 & -1.94 & 1, 20, 36, 42 &
NGC 3938 & 12.4 & 3.47 & \dots & 9.78 & -5.23 & 10, 36, 42 \\
NGC 4030 & 20.14 & 18.65 & 0.66 & 10.47 & 1.07 & 1, 20, 36, 42 &
NGC 4038 & 21.1 & 12.52 & 0.16 & 10.82 & 7.31 & 3, 21, 36, 44 \\
NGC 4041 & 21.57 & 8.31 & 0.23 & 10.38 & -0.67 & 3, 20, 36, 42 &
NGC 4150 & 4.11 & 0.01 & \dots & 7.85 & -6.4 & 15, 32, 41, 52 \\
NGC 4178 & 19.36 & 0.55 & \dots & 9.53 & -5.04 & 3, 37, 42 &
NGC 4189 & 20.06 & 2.46 & \dots & 9.71 & -5.54 & 3, 37, 42 \\
NGC 4254 & 16.3 & 18.39 & \dots & 10.59 & 6.96 & 4, 36, 42 &
NGC 4258 & 6.85 & 3.08 & \dots & 9.67 & -3.64 & 10, 39, 42 \\
NGC 4294 & 19.19 & 0.36 & \dots & 9.59 & -5.47 & 3, 37, 42 &
NGC 4299 & 15.9 & 0.25 & \dots & 9.4 & -5.88 & 3, 37, 42 \\
NGC 4303 & 19.8 & 18.41 & \dots & 10.73 & 14.43 & 4, 36, 42 &
NGC 4321 & 19.47 & 22.16 & \dots & 10.6 & 5.51 & 4, 36, 42 \\
NGC 4394 & 19.08 & 2.49 & \dots & 9.28 & -6.37 & 3, 37, 46 &
NGC 4402 & 18.6 & 4.4 & \dots & 9.94 & -5.29 & 4, 36, 42 \\
NGC 4414 & 8.48 & 5.19 & 0.11 & 9.91 & -4.34 & 2, 20, 36, 42 &
NGC 4459 & 19.53 & 0.52 & 0.06 & 9.48 & -6.32 & 15, 32, 40, 52 \\
NGC 4501 & 19.41 & 19.63 & \dots & 10.53 & 6.09 & 4, 36, 42 &
NGC 4519 & 19.67 & 1.14 & \dots & 9.71 & -5.58 & 3, 37, 42 \\
NGC 4526 & 18.79 & 1.55 & 0.12 & 9.88 & -5.93 & 15, 32, 36, 50 &
NGC 4535 & 19.75 & 13.83 & \dots & 10.26 & -2.78 & 4, 36, 42 \\
NGC 4548 & 18.79 & 6.13 & \dots & 9.49 & -6.21 & 4, 37, 42 &
NGC 4561 & 19.42 & 1.11 & \dots & 9.2 & -6.08 & 3, 37, 42 \\
NGC 4569 & 18.8 & 16.25 & 0.23 & 10.2 & -2.87 & 2, 22, 36, 42 &
NGC 4571 & 19.43 & 3.51 & \dots & 9.38 & -6.15 & 3, 37, 49 \\
NGC 4579 & 19.32 & 8.13 & \dots & 10.07 & -2.08 & 2, 36, 42 &
NGC 4631 & 7.35 & 2.17 & 0.06 & 10.18 & 1.36 & 7, 20, 36, 43 \\
NGC 4639 & 18.97 & 3.33 & \dots & 9.38 & -6.08 & 9, 37, 42 &
NGC 4647 & 19.1 & 5.53 & \dots & 10 & -4.03 & 4, 36, 42 \\
NGC 4651 & 19.12 & 3.13 & \dots & 9.91 & -4.94 & 3, 36, 42 &
NGC 4654 & 19.55 & 6.64 & \dots & 10.31 & -0.78 & 4, 36, 42 \\
NGC 4689 & 19.79 & 7.58 & \dots & 9.47 & -5.81 & 2, 37, 42 &
NGC 4698 & 19.51 & 0.84 & \dots & 9.07 & -6.34 & 3, 37, 46 \\
NGC 4710 & 19.23 & 2.2 & 0.11 & 9.88 & -5.74 & 16, 34, 36, 52 &
NGC 4713 & 19.43 & 0.65 & \dots & 9.73 & -4.4 & 3, 37, 42 \\
NGC 4736 & 4.37 & 1.3 & \dots & 9.64 & -5.79 & 2, 36, 42 &
NGC 4826 & 16.3 & 14.12 & 0.44 & 10.52 & -3.11 & 1, 20, 36, 42 \\
NGC 4945 & 5.2 & 15.74 & 0.47 & 10.73 & 7.19 & 9, 31, 36, 43 &
NGC 5005 & 14.1 & 12.36 & 0.62 & 10.24 & -1.89 & 1, 20, 36, 42 \\
NGC 5033 & 12.63 & 6.19 & 0.17 & 10.06 & -3 & 9, 24, 36, 42 &
NGC 5055 & 7.44 & 6.35 & 0.11 & 10.03 & -4.08 & 1, 20, 36, 42 \\
NGC 5194 & 7.53 & 14 & 0.06 & 10.29 & 3.71 & 10, 22, 36, 42 &
NGC 5236 & 4.7 & 7.07 & 0.06 & 10.33 & 0.06 & 1, 28, 36, 46 \\
NGC 5347 & 32.18 & 0.39 & 0.16 & 9.95 & -5.6 & 14, 24, 41, 44 &
NGC 5457 & 4.94 & 5.31 & \dots & 9.93 & -4.21 & 2, 36, 42 \\
NGC 5678 & 32.23 & 20.62 & 0.99 & 10.61 & 7.27 & 1, 20, 36, 42 &
NGC 5713 & 30.26 & 11.33 & 0.38 & 10.82 & 10.91 & 1, 20, 36, 42 \\
NGC 5775 & 27.61 & 16.2 & 0.84 & 10.82 & 16.49 & 1, 20, 36, 42 &
NGC 5861 & 28.5 & 4.13 & 0.12 & 10.49 & -3.24 & 13, 21, 36, 44 \\
NGC 5936 & 55.4 & 12.38 & 0.6 & 10.97 & 11.78 & 17, 35, 36, 50 &
NGC 6014 & 34.34 & 1.87 & 0.05 & 9.71 & -5.82 & 16, 34, 41, 48 \\
NGC 6207 & 17.55 & 0.75 & \dots & 9.68 & -5.11 & 3, 37, 42 &
NGC 6217 & 24.26 & 6.46 & \dots & 10.37 & -0.77 & 2, 36, 42 \\
NGC 6503 & 6.27 & 0.99 & \dots & 9.22 & -6.23 & 3, 36, 42 &
NGC 6643 & 25.79 & 8.78 & \dots & 10.51 & 1.14 & 6, 36, 42 \\
NGC 6814 & 21.5 & 1.8 & 0.18 & 10.15 & -3.65 & 14, 24, 36, 44 &
NGC 6946 & 4.16 & 5.16 & 0.23 & 9.94 & -3.5 & 2, 20, 36, 42 \\
NGC 6951 & 23.01 & 18.86 & 0.34 & 10.53 & -2.09 & 2, 22, 36, 44 &
NGC 7252 & 64.9 & 9.27 & \dots & 10.73 & 6.4 & 3, 37, 48 \\
NGC 7331 & 14.38 & 27.11 & 0.37 & 10.56 & 1.74 & 6, 20, 36, 42 &
NGC 7465 & 27.90 & 2.46 & 0.02 & 10.11 & -4.62 & 16, 34, 36, 48 \\
NGC 7479 & 33.56 & 23.08 & 0.94 & 10.82 & 8.02 & 9, 20, 36, 42 &
NGC 7582 & 21.66 & 7.98 & 0.41 & 10.88 & 8.76 & 13, 21, 36, 46 \\
UGC 4013 & 122.84 & 4.38 & 2.31 & 10.78 & 4.07 & 14, 24, 41, 44 &
He 2-10 & 12 & 2.1 & 0.04 & 10.04 & -4.96 & 19, 33, 40, 44 \\
IC 342 & 4.32 & 16.32 & 0.4 & 10.11 & -1.38 & 2, 20, 36, 42 &
IC 676 & 21 & 1.09 & 0.02 & 9.69 & -5.85 & 16, 34, 41, 48 \\
Maffei 2 & 4.1 & 6.72 & 0.09 & 9.73 & -5.82 & 2, 30, 40, 44 &
& & & & & & \\
\cline{1-14}  \\
 &  & \multicolumn{10}{c}{\bf (U)LIRGs} & &  \\
\cline{1-14}  \\
Arp 055 & 165.58 & 121.58 & 4.31 & 11.68 & 114.32 & 1, 20, 36, 42 &
Arp 148 & 147.13 & 45.54 & 4.84 & 11.6 & 56.54 & 1, 20, 36, 42 \\
Arp 193 & 101.84 & 48.55 & 2.83 & 11.66 & 122.66 & 5, 23, 36, 42 &
Arp 220 & 80.89 & 90.59 & 10.78 & 12.2 & 248.83 & 3, 29, 36, 42 \\
IC 0883 & 101.84 & 55.81 & \dots & 11.66 & 122.66 & 3, 36, 42 &
IC 1623 & 80.72 & 118.73 & 7.36 & 11.65 & 158.1 & 1, 20, 36, 44 \\
IC 5179 & 44.65 & 21.78 & 3.06 & 11.11 & 32.96 & 1, 20, 36, 45 &
III Zw 35 & 110 & 18.55 & 2.7 & 11.57 & 52.53 & 17, 35, 36, 48 \\
00335-2732 & 281.69 & 52.01 & 50.59 & 11.94 & 100.89 & 13, 21, 41, 44 &
00506+7248 & 65.2 & 22.81 & 0.78 & 11.44 & 56.67 & 17, 35, 36, 48 \\
01364-1042 & 205.92 & 48.23 & 12.11 & 11.83 & 79.85 & 13, 21, 36, 50 &
03056+2034 & 114.09 & 30.09 & 1.02 & 11.25 & 20.23 & 13, 21, 40, 44 \\
05189-2524 & 173.26 & 67.85 & 6.59 & 12.1 & 98.12 & 1, 20, 36, 44 &
05414+5840 & 61.63 & 38.19 & 2.34 & 11.22 & 39.96 & 13, 21, 36, 44 \\
07329+1149 & 69.15 & 9.9 & 0.51 & 11.08 & 20.49 & 17, 35, 36, 48 &
08071+0509 & 222.9 & 86.16 & 7.82 & 11.84 & 160.05 & 13, 21, 41, 49 \\
10039-3338 & 143.7 & 32.83 & 7.69 & 11.71 & 51.42 & 18, 21, 36, 44 &
10173+0828 & 204.09 & 54.36 & 3.71 & 11.78 & 47.43 & 5, 35, 36, 44 \\
10565+2448 & 173.3 & 56.93 & 10.69 & 11.97 & 201.3 & 1, 20, 36, 48 &
11506-3851 & 40.26 & 8.93 & 2.78 & 11.15 & 12.22 & 13, 21, 36, 49 \\
12112+0305 & 313.55 & 114.95 & 14.65 & 12.29 & 283.45 & 13, 27, 36, 52 &
13126+2452 & 54.08 & 16.9 & 0.34 & 11.09 & 4.27 & 13, 21, 36, 50 \\
16399-0937 & 113.18 & 37 & 1.43 & 11.52 & 77.44 & 13, 21, 36, 44 &
17138-1017 & 75.0 & 22.62 & 0.67 & 11.39 & 40.24 & 17, 35, 41, 48 \\
17208-0014 & 175.81 & 115.15 & 16.38 & 12.35 & 348.64 & 5, 21, 36, 47 &
17526+3253 & 103.36 & 27.18 & 1.59 & 11.08 & 21.72 & 13, 21, 41, 49 \\
18090+0130 & 121.3 & 41.23 & 1.55 & 11.56 & 96.41 & 17, 35, 36, 48 &
18293-3413 & 79.89 & 92.39 & 4.94 & 11.81 & 166.95 & 1, 20, 36, 48 \\
20550+1656 & 147.84 & 42.93 & 2.53 & 11.87 & 106.58 & 13, 21, 36, 44 &
22025+4205 & 60.68 & 8.09 & 2.07 & 10.97 & 8.93 & 13, 21, 36, 44 \\
23365+3604 & 258.74 & 83.84 & 16.35 & 12.11 & 224.29 & 1, 20, 36, 48 &
Mrk 0231 & 173.37 & 62.08 & 20.78 & 12.48 & 863.92 & 5, 27, 36, 42 \\
Mrk 0266 & 121.77 & 76.16 & 1.29 & 11.83 & 192.31 & 3, 26, 37, 52 &
Mrk 0273 & 159.66 & 66.72 & 14.35 & 12.14 & 434.95 & 1, 27, 36, 42 \\
Mrk 0331 & 71.92 & 42.55 & 2.76 & 11.41 & 37.36 & 1, 20, 36, 44 &
Mrk 1027 & 121.3 & 48.95 & 2.68 & 11.38 & 87.26 & 1, 20, 36, 42 \\
NGC 0023 & 62.4 & 16.23 & 0.65 & 11.08 & 23.88 & 17, 35, 36, 50 &
NGC 0034 & 80.92 & 35.32 & 8.97 & 11.46 & 39.59 & 14, 24, 36, 50 \\
NGC 0695 & 130.6 & 89.49 & 4.73 & 11.63 & 146.25 & 1, 20, 36, 42 &
NGC 0828 & 72.73 & 78.92 & 0.81 & 11.32 & 59.42 & 3, 29, 36, 42 \\
NGC 1068 & 15.46 & 29.56 & 2.99 & 11.37 & 132.3 & 9, 20, 36, 42 &
NGC 1144 & 115.93 & 71.3 & 1.95 & 11.39 & 242.85 & 1, 20, 37, 42 \\
NGC 1365 & 21.84 & 43.28 & 3.55 & 11.17 & 15.11 & 1, 20, 36, 44 &
NGC 1614 & 64.78 & 32.86 & 1.25 & 11.61 & 62.99 & 12, 20, 36, 44 \\
NGC 2146 & 15.34 & 14.34 & 1.03 & 11 & 23.85 & 1, 20, 36, 44 &
NGC 2369 & 44.72 & 27.09 & 1.97 & 11.1 & 28.54 & 13, 21, 36, 51 \\
NGC 2388 & 60.4 & 22.85 & 1.14 & 11.25 & 21.8 & 17, 35, 36, 52 &
NGC 2623 & 76.71 & 24.47 & 1.41 & 11.52 & 60.98 & 3, 26, 36, 42 \\
NGC 3034 & 4.89 & 10.34 & 0.45 & 11.03 & 15.51 & 1, 20, 36, 42 &
NGC 3110 & 73.24 & 24.74 & 1.02 & 11.29 & 76.4 & 17, 35, 36, 48 \\
NGC 3256 & 35.86 & 46.43 & 2.48 & 11.56 & 95 & 11, 21, 36, 46 &
NGC 3690 & 48.02 & 34.41 & 2.62 & 11.88 & 175.15 & 3, 29, 36, 47 \\
NGC 5135 & 54.08 & 30.5 & 2.57 & 11.19 & 61.49 & 1, 20, 36, 45 &
NGC 5653 & 55.31 & 16.44 & 0.63 & 11.05 & 20.39 & 17, 35, 36, 48 \\
NGC 6240 & 106.71 & 83.26 & 14.71 & 11.85 & 527.69 & 5, 21, 36, 45 &
NGC 6701 & 59.07 & 32.51 & 1.3 & 11.07 & 30.72 & 1, 20, 36, 42 \\
NGC 6921 & 59.41 & 30.22 & 2.41 & 11.07 & 5.17 & 3, 20, 40, 42 &
NGC 7130 & 68.58 & 44.06 & 3.21 & 11.37 & 96.58 & 1, 20, 36, 45 \\
NGC 7469 & 68.08 & 33.31 & 2.01 & 11.61 & 93.98 & 1, 20, 36, 42 &
NGC 7552 & 22.05 & 11 & 0.69 & 11.05 & 9.66 & 11, 21, 36, 45 \\
NGC 7591 & 67.37 & 16.45 & 0.57 & 11.01 & 15.87 & 17, 35, 41, 50 &
NGC 7771 & 58.04 & 30.46 & 3.99 & 11.34 & 50.43 & 12, 23, 36, 42 \\
UGC 1845 & 63.96 & 19.79 & 1.09 & 11.08 & 24.14 & 17, 35, 36, 48 &
UGC 5101 & 164.93 & 48.77 & 11.71 & 11.95 & 546.91 & 1, 20, 36, 42 \\
VII Zw 31 & 220.72 & 226.41 & 10.69 & 11.92 & 237.84 & 3, 20, 36, 44 &
ZW 049.057 & 53.38 & 8.48 & 1.53 & 11.18 & 10.99 & 5, 21, 36, 47 \\
 
\enddata

\tablerefs{\mbox{{\bf CO data:} [1] \citet{gao2004a}; [2] \citet{kuno2007}; [3] \citet{young1995}; 
	[4] \citet{chung-E2009}; [5] \citet{papadopoulos2010};} 
	\mbox{[6] \citet{nishiyama2001}; [7] \citet{paglione2001}; [8] \citet{sage1993}; [9] \citet{maiolino1997};
	[10] \citet{helfer2003}; [11] \citet{smith1996}; [12] \citet{elfhag1996};}
	\mbox{[13] \citet{baan2008}; [14] \citet{curran2000}; [15] \citet{young2008}; [16] \citet{crocker2012};
        [17] \citet{GB-2012}; [18] \citet{gao1999};}
        \mbox{[19] \citet{bayet2009}. {\bf HCN data:} [20] \citet{gao2004a}; [21] \citet{baan2008}; [22] \citet{krips2008};
        [23] \citet{juneau2009}; [24] \citet{curran2000};} 
        \mbox{[25] \citet{matsushita2010}; [26] \citet{imanishi2009}; [27] \citet{GC-A-2008}; [28] \citet{huettemeister1995};
	[29] \citet{solomon1992}; [30] \citet{nguyen1992};} 
	\mbox{[31] \citet{wang2004}; [32] \citet{krips2010}; [33] \citet{imanishi2007}; [34] \citet{crocker2012}; 
	[35] \citet{GB-2012}. {\bf IR data:} [36] \citet{sanders2003};} 
	\mbox{[37] \citet{moshir1990}; [38] \citet{lisenfeld2007}; [39] \citet{rice1988}; [40] \citet{ipac1986};
	[41] \citet{moshir1993}. {\bf 1.4 GHz RC data:} [42] \citet{condon2002};} 
	\mbox{[43] \citet{strickland2004}; [44] \citet{condon1998}; [45] \citet{condon1996}; [46] \citet{condon1987}; 
	[47] \citet{baan2006}; [48] \citet{yun2001}; [49] measured from our}
	\mbox{ radio maps;  [50] \citet{condon1990}; [51] \citet{mauch2008}; [52] \citet{becker2003}.}
}

\tablecomments{\mbox{D is the distance in [Mpc]; ${L_{\rm CO}}$ and ${L_{\rm HCN}}$ are line luminosities in [$10^8 
	~{\rm K~km^{-1}~pc^{2}}$]; ${L_{\rm IR}}$ is the log of total IR luminosity ($8-1000{\rm \mu m}$) in 
        [${\rm L_\odot}$]; and ${L_{\rm RC}}$ is the}
        \mbox{luminosity of 1.4 GHz RC in [$10^{21}~{\rm W~Hz^{-1}}$].}
		}

\end{deluxetable*}

\clearpage
%%%%%%%%%%%%%%%%%%%%%%%%%%%%%%%%%%%%%%%
%                 Table 2
%%%%%%%%%%%%%%%%%%%%%%%%%%%%%%%%%%%%%%%
\LongTables
\begin{deluxetable*}{lccccccccc}
\tabletypesize{\tiny}
\renewcommand{\arraystretch}{2.0}
\setlength{\tabcolsep}{5pt}
\tablecolumns{10}
\tablewidth{1pc}
\tablecaption{Surface Density \label{table:table2}}
\tablehead{
\colhead{\bf Name} & \colhead{\bf Diameter} & \colhead{\bf log $\Sigma _ {\rm {H\,{\scriptsize I}}}$} & \colhead{\bf log\tablenotemark{a}$\Sigma _ {\rm {H2}}$}
& \colhead{\bf log\tablenotemark{b}$\Sigma _ {\rm {H2}}$} & \colhead{\bf log\tablenotemark{a}$\Sigma _ {\rm {gas}}$} & \colhead{\bf log $\Sigma _ {\rm {den}}$}
& \colhead{\bf log $\Sigma _ {\rm {IR}}$} & \colhead{\bf log $\Sigma _ {\rm {RC}}$} & \colhead{Ref.} \\
\cline{3-7} \\
\colhead{} & [arcsec] &\multicolumn{5}{c}{${\rm [M_\odot~pc^{-2}]}$}&\multicolumn{2}{c}{${\rm [M_\odot~yr^{-1}~kpc^{-2}]}$}& \\
}

\startdata
 &  & \multicolumn{6}{c}{\bf Normal Galaxies} & &  \\
\cline{1-10}  \\
NGC 0134 & 150 & 0.29 & 0.66 & 0.93 & 0.81 & 0.02 & -1.83 & -1.84 & 4, 33 \\
NGC 0174 & 5.1 & \dots & 3.39 & 3.02 & 3.39 & 2.88 & 0.67 & 0.37 & AU23,  \\
NGC 0253 & 96 & 0.82 & 3.09 & 2.81 & 3.09 & 2.06 & 0.05 & 0.04 & 1, 24 \\
NGC 0520 & 4.5 & 0.73 & 4.28 & 3.62 & 4.28 & 2.94 & 1.19 & 1.19 & AC205, 27 \\
NGC 0598 & 1713 & 0.89 & -0.39 & -0.12 & 0.91 & \dots & -2.45 & -2.34 & 2, 18 \\
NGC 0628 & 329 & 0.62 & 0.86 & 1.13 & 1.06 & \dots & -2.47 & -2.41 & 1, 13 \\
NGC 0660 & 18 & 0.19 & 3.43 & 3.04 & 3.43 & 1.99 & 0.37 & 0.34 & AL296, 17 \\
NGC 0772 & 79 & 0.48 & 1.9 & 2 & 1.92 & \dots & -1.7 & -1.69 & 1, 19 \\
NGC 0891 & 238 & 0.21 & 1.3 & 1.57 & 1.33 & 0.1 & -1.76 & -1.65 & 4, 17 \\
NGC 0925 & 186 & 0.85 & 0.93 & 1.2 & 1.19 & \dots & -2.35 & -2.33 & 1, 13 \\
NGC 0931 & 50 & \dots & -0.14 & 0.13 & -0.14 & 0.65 & -1.66 & -2.02 & 9,  \\
NGC 1022 & 5.0 & 0.4 & 3.61 & 3.17 & 3.61 & 2.62 & 0.96 & 0.67 & AG278+AL377, 20 \\
NGC 1055 & 90 & 0.72 & 1.92 & 2.02 & 1.95 & 0.64 & -1.33 & -1.3 & 1, 17 \\
NGC 1058 & 90 & 0.5 & 0.78 & 1.05 & 0.96 & \dots & -2.04 & -2.21 & AP404, 15 \\
NGC 1097 & 28 & 0.14 & 2.42 & 2.36 & 2.42 & \dots & -0.08 & -0.02 & AB775, 28 \\
NGC 1266 & 5.1 & \dots & 3.87 & 3.34 & 3.87 & 2.78 & 0.69 & 0.89 & 11,  \\
NGC 1530 & 31 & 0.77 & 2.18 & 2.2 & 2.2 & 1.04 & -0.85 & -0.9 & 4, 15 \\
NGC 1569 & 74 & 1.26 & 0.52 & 0.79 & 1.33 & \dots & -0.86 & -0.68 & 1, 16 \\
NGC 1667 & 33 & \dots & 1.7 & 1.87 & 1.7 & 1.65 & -1.11 & -0.9 & AA281,  \\
NGC 1808 & 17 & 0.72 & 3.26 & 2.93 & 3.26 & 2.62 & 0.61 & 0.51 & AD231, 29 \\
NGC 2273 & 1.7 & 0.3 & 3.6 & 3.16 & 3.6 & 3.38 & 1.47 & 1.55 & AY044, 31 \\
NGC 2276 & 76 & 1.02 & 1.19 & 1.46 & 1.41 & 0.12 & -1.51 & -1.04 & 1, 15 \\
NGC 2403 & 328 & 0.85 & 0.18 & 0.45 & 0.93 & \dots & -2.15 & -2.15 & 1, 13 \\
NGC 2764 & 9.4 & 0.93 & 2.71 & 2.55 & 2.72 & 1.25 & -0.26 & -0.48 & 11, 20 \\
NGC 2841 & 216 & 0.44 & 1.09 & 1.36 & 1.18 & \dots & -2.55 & -2.35 & 1, 13 \\
NGC 2903 & 122 & 0.74 & 1.77 & 1.92 & 1.81 & 0.56 & -1.22 & -1.21 & 1, 13 \\
NGC 2976 & 154 & 0.9 & 0.89 & 1.16 & 1.2 & \dots & -1.85 & -1.97 & 1, 13 \\
NGC 3031 & 466 & 0.35 & 0.2 & 0.47 & 0.58 & \dots & -2.31 & -2.27 & 1, 18 \\
NGC 3032 & 15 & -0.16 & 2.1 & 2.14 & 2.1 & 0.9 & -0.81 & -1 & 12, 38 \\
NGC 3079 & 38 & 0.44 & 2.83 & 2.64 & 2.83 & 1.78 & -0.4 & 0.01 & ASTUD, 30 \\
NGC 3147 & 60 & 0.28 & 2.06 & 2.11 & 2.07 & 0.42 & -1.35 & -1.3 & 1, 15 \\
NGC 3310 & 32 & 0.97 & 1.62 & 1.81 & 1.71 & \dots & -0.43 & -0.45 & AD176, 15 \\
NGC 3338 & 145 & 0.59 & 0.44 & 0.71 & 0.82 & \dots & -2.55 & -2.42 & 1, 19 \\
NGC 3351 & 16 & 0.28 & 3.25 & 2.93 & 3.25 & \dots & 0.08 & -0.22 & AO73, 13 \\
NGC 3368 & 97 & 0.58 & 1.3 & 1.57 & 1.38 & \dots & -1.78 & -1.91 & 1, 18 \\
NGC 3486 & 129 & 0.91 & 0.94 & 1.21 & 1.23 & \dots & -2.09 & -1.94 & 1, 18 \\
NGC 3504 & 8.0 & 0.43 & 3.48 & 3.08 & 3.48 & 2.59 & 0.6 & 0.88 & AS204, 31 \\
NGC 3521 & 166 & 1.03 & 1.94 & 2.03 & 1.99 & \dots & -1.55 & -1.54 & 1, 13 \\
NGC 3556 & 185 & 1.12 & 0.96 & 1.23 & 1.35 & -0.26 & -1.86 & -1.8 & 1, 15 \\
NGC 3620 & 26 & \dots & 2.41 & 2.35 & 2.41 & 1.86 & -0.16 & -0.4 & 10,  \\
NGC 3627 & 160 & 0.73 & 1.73 & 1.89 & 1.77 & -0.03 & -1.43 & -1.41 & 1, 13 \\
NGC 3628 & 71 & 0.85 & 2.42 & 2.36 & 2.43 & 1.3 & -0.85 & -0.73 & 1, 21 \\
NGC 3631 & 85 & \dots & 1.67 & 1.85 & 1.67 & \dots & -1.64 & -1.63 & 1,  \\
NGC 3665 & 31 & \dots & 1.44 & 1.69 & 1.44 & 0.46 & -1.48 & -0.86 & 11,  \\
NGC 3675 & 89 & 0.56 & 1.74 & 1.9 & 1.77 & \dots & -1.55 & -1.74 & 1, 18 \\
NGC 3726 & 174 & 0.8 & 0.86 & 1.13 & 1.13 & \dots & -2.37 & -2.43 & 1, 15 \\
NGC 3893 & 81 & 1.19 & 1.46 & 1.71 & 1.65 & 0.65 & -1.45 & -1.38 & 1, 15 \\
NGC 3938 & 149 & 0.82 & 1.1 & 1.37 & 1.28 & \dots & -2.1 & -2.1 & 1, 18 \\
NGC 4030 & 74 & 0.55 & 2.02 & 2.09 & 2.03 & 0.91 & -1.31 & -1.3 & 1, 22 \\
NGC 4038 & 49 & \dots & 2.16 & 2.18 & 2.16 & 0.6 & -0.66 & -0.72 & 9,  \\
NGC 4041 & 35 & \dots & 2.26 & 2.25 & 2.26 & 1.03 & -0.8 & -0.81 & AC101,  \\
NGC 4150 & 17 & -0.31 & 1.74 & 1.9 & 1.74 & 0.59 & -0.65 & -1.21 & 11, 15 \\
NGC 4178 & 166 & 0.93 & -0.18 & 0.09 & 0.96 & \dots & -2.8 & -2.53 & 1, 18 \\
NGC 4189 & 73 & 0.88 & 1.16 & 1.43 & 1.34 & \dots & -1.96 & -2 & 9, 14 \\
NGC 4254 & 137 & 0.96 & 1.66 & 1.84 & 1.74 & \dots & -1.54 & -1.4 & 1, 14 \\
NGC 4258 & 207 & 0.57 & 1.28 & 1.55 & 1.36 & \dots & -1.97 & -1.59 & BG062, 18 \\
NGC 4294 & 56 & 1.3 & 0.59 & 0.86 & 1.38 & \dots & -1.8 & -1.7 & AV298, 14 \\
NGC 4299 & 45 & 1.29 & 0.78 & 1.05 & 1.41 & \dots & -1.6 & -1.54 & AV298, 14 \\
NGC 4303 & 136 & 0.84 & 1.5 & 1.73 & 1.59 & \dots & -1.57 & -1.37 & 1, 18 \\
NGC 4321 & 123 & 0.51 & 1.68 & 1.86 & 1.71 & \dots & -1.6 & -1.51 & 1, 14 \\
NGC 4394 & 75 & -0.6 & 1.18 & 1.45 & 1.19 & \dots & -2.3 & -3.11 & 1, 14 \\
NGC 4402 & 54 & 0.87 & 1.74 & 1.89 & 1.79 & \dots & -1.43 & -1.58 & 5, 14 \\
NGC 4414 & 71 & 0.51 & 2.25 & 2.24 & 2.26 & 0.93 & -1.02 & -0.94 & 1, 19 \\
NGC 4459 & 6.5 & \dots & 2.61 & 2.48 & 2.61 & 2.02 & -0.03 & -0.65 & 12,  \\
NGC 4501 & 121 & 0.6 & 1.65 & 1.83 & 1.69 & \dots & -1.64 & -1.47 & 1, 14 \\
NGC 4519 & 88 & \dots & 0.67 & 0.94 & 0.67 & \dots & -2.1 & -2.16 & 1,  \\
NGC 4526 & 11 & 0.71 & 2.66 & 2.52 & 2.66 & 1.87 & -0.12 & -0.49 & 5, 39 \\
NGC 4535 & 133 & 0.57 & 1.4 & 1.66 & 1.46 & \dots & -1.99 & -2.04 & 1, 14 \\
NGC 4548 & 96 & -0.26 & 1.37 & 1.64 & 1.38 & \dots & -2.32 & -2.67 & AB342+AC101, 14 \\
NGC 4561 & 57 & 1.2 & 1.05 & 1.32 & 1.43 & \dots & -2.14 & -2.07 & 9, 14 \\
NGC 4569 & 85 & 0.61 & 1.9 & 2 & 1.92 & 0.39 & -1.61 & -1.62 & 1, 14 \\
NGC 4571 & 94 & \dots & 1.12 & 1.39 & 1.12 & \dots & -2.43 & -2.59 & 1,  \\
NGC 4579 & 57 & 0.1 & 1.92 & 2.02 & 1.93 & \dots & -1.4 & -1.23 & 1, 14 \\
NGC 4631 & 298 & 0.37 & 0.75 & 1.02 & 0.9 & -0.47 & -1.9 & -1.62 & 1, 17 \\
NGC 4639 & 83 & 0.64 & 1.22 & 1.49 & 1.32 & \dots & -2.3 & -2.38 & AK205, 18 \\
NGC 4647 & 74 & 0.59 & 1.54 & 1.76 & 1.59 & \dots & -1.68 & -1.64 & 9, 18 \\
NGC 4651 & 54 & 0.91 & 1.56 & 1.78 & 1.65 & \dots & -1.48 & -1.52 & 1, 14 \\
NGC 4654 & 108 & 0.93 & 1.27 & 1.54 & 1.43 & \dots & -1.75 & -1.71 & 1, 14 \\
NGC 4689 & 68 & 0.49 & 1.72 & 1.88 & 1.74 & \dots & -2.08 & -2.05 & 1, 14 \\
NGC 4698 & 81 & 0.27 & 0.62 & 0.89 & 0.78 & \dots & -2.55 & -2.96 & AC101+AB342, 14 \\
NGC 4710 & 11 & \dots & 2.79 & 2.61 & 2.79 & 1.81 & -0.13 & -0.4 & 5,  \\
NGC 4713 & 66 & 1.17 & 0.69 & 0.96 & 1.29 & \dots & -1.83 & -1.61 & AV298, 14 \\
NGC 4736 & 104 & 0.91 & 1.89 & 2 & 1.93 & \dots & -1.01 & -1.09 & 1, 13 \\
NGC 4826 & 39 & 0.88 & 2.64 & 2.51 & 2.65 & 1.47 & -0.51 & -0.84 & AS541, 13 \\
NGC 4945 & 47 & -0.02 & 3.52 & 3.1 & 3.52 & 2.33 & 0.51 & 0.53 & 10, 26 \\
NGC 5005 & 69 & 0.11 & 2.21 & 2.22 & 2.21 & 1.25 & -1.15 & -1.11 & 1, 22 \\
NGC 5033 & 63 & 0.79 & 2.09 & 2.13 & 2.11 & 0.85 & -1.13 & -1.02 & 1, 18 \\
NGC 5055 & 185 & 0.7 & 1.62 & 1.81 & 1.67 & 0.18 & -1.63 & -1.62 & 1, 13 \\
NGC 5194 & 284 & 0.58 & 1.58 & 1.79 & 1.62 & -0.44 & -1.78 & -1.48 & 1, 13 \\
NGC 5236 & 321 & 0.44 & 1.59 & 1.79 & 1.62 & -0.13 & -1.44 & -1.37 & 1, 13 \\
NGC 5347 & 36 & 0.58 & 0.56 & 0.83 & 0.87 & 0.51 & -1.55 & -1.81 & 9, 15 \\
NGC 5457 & 452 & 0.58 & 1.12 & 1.39 & 1.23 & \dots & -2.14 & -2.06 & AC101+AF085, 13 \\
NGC 5678 & 57 & 1.02 & 1.88 & 1.99 & 1.94 & 0.9 & -1.36 & -1.22 & 9, 19 \\
NGC 5713 & 33 & 0.96 & 2.15 & 2.17 & 2.18 & 1.01 & -0.63 & -0.59 & AL377, 23 \\
NGC 5775 & 115 & 0.62 & 1.3 & 1.57 & 1.38 & 0.35 & -1.63 & -1.47 & 6, 24 \\
NGC 5861 & 73 & \dots & 1.08 & 1.35 & 1.08 & -0.13 & -1.57 & -1.88 & 9,  \\
NGC 5936 & 9.0 & \dots & 2.79 & 2.61 & 2.79 & 1.81 & 0.12 & 0.04 & 5,  \\
NGC 6014 & 4.4 & \dots & 3.01 & 2.76 & 3.01 & 1.76 & 0.01 & -0.15 & 11,  \\
NGC 6207 & 76 & 0.82 & 0.72 & 0.99 & 1.07 & \dots & -1.91 & -1.78 & 9, 19 \\
NGC 6217 & 16 & \dots & 2.73 & 2.57 & 2.73 & \dots & -0.23 & -0.24 & AC205+AC101,  \\
NGC 6503 & 118 & 0.68 & 1.35 & 1.62 & 1.43 & \dots & -1.77 & -1.93 & 1, 19 \\
NGC 6643 & 74 & 1.08 & 1.48 & 1.72 & 1.63 & \dots & -1.48 & -1.51 & 4, 15 \\
NGC 6814 & 93 & 0.73 & 0.75 & 1.02 & 1.04 & 0.08 & -1.85 & -1.89 & 9, 36 \\
NGC 6946 & 318 & 0.87 & 1.56 & 1.78 & 1.64 & 0.56 & -1.68 & -1.51 & 1, 13 \\
NGC 6951 & 82 & 0.27 & 1.82 & 1.95 & 1.83 & 0.41 & -1.46 & -1.7 & 9, 35 \\
NGC 7252 & 5.0 & \dots & 3.04 & 2.78 & 3.04 & \dots & 0.26 & 0.26 & AN90,  \\
NGC 7331 & 146 & 0.91 & 1.88 & 1.99 & 1.92 & 0.35 & -1.52 & -1.56 & 1, 13 \\
NGC 7465 & 7.5 & 0.86 & 2.84 & 2.65 & 2.84 & 1.11 & 0.07 & -0.07 & 11, 40 \\
NGC 7479 & 97.9 & 0.79 & 1.43 & 1.68 & 1.52 & 0.37 & -1.67 & -1.7 & 1, 25 \\
NGC 7582 & 35 & 0.31 & 2.24 & 2.23 & 2.25 & 1.29 & -0.33 & -0.4 & 4, 34 \\
UGC 4013 & 32 & \dots & 0.55 & 0.82 & 0.55 & 0.61 & -1.86 & -1.99 & AC345,  \\
He 2-10 & 9.1 & 0.93 & 3.34 & 2.98 & 3.34 & 1.92 & 0.58 & 0.43 & AH173, 37 \\
IC 342 & 529 & 0.32 & 1.59 & 1.79 & 1.61 & 0.32 & -2 & -1.81 & 1, 17 \\
IC 676 & 6.2 & \dots & 2.9 & 2.69 & 2.9 & 1.59 & 0.12 & -0.04 & 11,  \\
Maffei 2 & 23.3 & 0.54 & 3.96 & 3.41 & 3.96 & 2.41 & 0.42 & 0.24 & AH407, 32 \\
\cline{1-10}  \\
 &  & \multicolumn{6}{c}{\bf (U)LIRGs} & &  \\
\cline{1-10}  \\
Arp 055 & 2.2+0.8 & \dots & 4 & 3.43 & 4 & 2.89 & 1.15 & 1.08 & 5,  \\
Arp 148 & 0.24 & \dots & 5.66 & 4.56 & 5.66 & 5.02 & 3.16 & 2.88 & 7,  \\
Arp 193 & 2.9 & \dots & 3.84 & 3.32 & 3.84 & 2.94 & 1.37 & 1.34 & 5,  \\
Arp 220 & 0.34+0.22 & \dots & 6.02 & 4.81 & 6.02 & 5.43 & 3.81 & 3.55 & 7,  \\
IC 0883 & 2.2 & \dots & 4.14 & 3.53 & 4.14 & \dots & 1.61 & 1.58 & AB370,  \\
IC 1623 & 15 & \dots & 3 & 2.75 & 3 & 2.13 & 0.14 & 0.22 & AS267,  \\
IC 5179 & 33 & \dots & 2.1 & 2.14 & 2.1 & 1.58 & -0.56 & -0.57 & 4,  \\
III Zw 35 & 0.18 & \dots & 5.77 & 4.64 & 5.77 & 5.27 & 3.63 & 3.35 & 7,  \\
IR 00335-2732 & 0.36 & \dots & 4.8 & 3.98 & 4.8 & 5.12 & 2.57 & 2.19 & AN054,  \\
IR 00506+7248 & 3.2 & \dots & 3.41 & 3.03 & 3.41 & 2.28 & 1.05 & 0.93 & 11,  \\
IR 01364-1042 & 0.19 & \dots & 5.59 & 4.52 & 5.59 & 5.33 & 3.29 & 2.93 & 7,  \\
IR 03056+2034 & 0.47 & \dots & 5.12 & 4.19 & 5.12 & 3.98 & 2.45 & 2.14 & 11,  \\
IR 05189-2524 & 0.22 & \dots & 5.76 & 4.63 & 5.76 & 5.09 & 3.57 & 3.03 & 7,  \\
IR 05414+5840 & 4.9 & \dots & 3.72 & 3.24 & 3.72 & 2.84 & 0.92 & 0.88 & AS192+AM293,  \\
IR 07329+1149 & 5.1 & \dots & 3.4 & 3.02 & 3.4 & 2.45 & 1.06 & 0.91 & AC685+AC345,  \\
IR 08071+0509 & 1.2 & \dots & 4.18 & 3.55 & 4.18 & 3.47 & 1.63 & 1.54 & AB660,  \\
IR 10039-3338 & 0.72 & \dots & 4.58 & 3.83 & 4.58 & 4.29 & 2.33 & 1.91 & 11,  \\
IR 10173+0828 & 0.13 & \dots & 5.98 & 4.78 & 5.98 & 5.15 & 3.58 & 3.06 & AS530,  \\
IR 10565+2448 & 0.80 & \dots & 4.57 & 3.82 & 4.57 & 4.18 & 2.33 & 2.21 & 7,  \\
IR 11506-3851 & 3.8 & \dots & 3.68 & 3.21 & 3.68 & 3.51 & 1.44 & 1.08 & AM93,  \\
IR 12112+0305 & 0.16+0.12 & \dots & 5.56 & 4.49 & 5.56 & 5 & 3.33 & 3.04 & AN122,  \\
IR 13126+2452 & 0.36 & \dots & 5.74 & 4.62 & 5.74 & 4.38 & 3.18 & 2.62 & 11,  \\
IR 16399-0937 & 10.8+5.1 & \dots & 2.4 & 2.35 & 2.4 & 1.33 & -0.09 & -0.16 & AB660,  \\
IR 17138-1017 & 5.9 & \dots & 3.16 & 2.86 & 3.16 & 1.97 & 0.75 & 0.55 & 4,  \\
IR 17208-0014 & 0.58 & \dots & 5.14 & 4.21 & 5.14 & 4.63 & 2.97 & 2.71 & AH640,  \\
IR 17526+3253 & 1.6+1.3 & \dots & 3.87 & 3.35 & 3.87 & 2.98 & 1.08 & 0.97 & AM293,  \\
IR 18090+0130 & 3.0 & \dots & 3.59 & 3.15 & 3.59 & 2.5 & 1.09 & 1.06 & AB275,  \\
IR 18293-3413 & 12 & \dots & 3.1 & 2.82 & 3.1 & 2.16 & 0.49 & 0.45 & 4,  \\
IR 20550+1656 & 7.3+5.1 & \dots & 2.49 & 2.41 & 2.49 & 1.6 & 0.27 & -0.01 & AT149,  \\
IR 22025+4205 & 1.0 & \dots & 4.44 & 3.73 & 4.44 & 4.18 & 1.95 & 1.79 & AM293,  \\
IR 23365+3604 & 0.21 & \dots & 5.55 & 4.49 & 5.55 & 5.18 & 3.28 & 3.07 & AB660,  \\
Mrk 0231 & 0.44 & \dots & 5.12 & 4.2 & 5.12 & 4.99 & 3.35 & 3.35 & 5,  \\
Mrk 0266 & 13 & \dots & 2.58 & 2.47 & 2.58 & 1.14 & 0.07 & 0.07 & 5,  \\
Mrk 0273 & 0.32+0.17 & \dots & 5.4 & 4.38 & 5.4 & 5.07 & 3.25 & 3.29 & 7,  \\
Mrk 0331 & 2.53 & \dots & 4.2 & 3.57 & 4.2 & 3.35 & 1.55 & 1.3 & AC243,  \\
Mrk 1027 & 11 & \dots & 2.53 & 2.44 & 2.53 & 1.61 & -0.22 & -0.1 & 5,  \\
NGC 0023 & 8.0 & \dots & 2.91 & 2.69 & 2.91 & 1.85 & 0.35 & 0.26 & AD255,  \\
NGC 0034 & 0.4 & \dots & 5.62 & 4.54 & 5.62 & 5.37 & 3.09 & 2.82 & 7,  \\
NGC 0695 & 13 & \dots & 2.59 & 2.47 & 2.59 & 1.65 & -0.18 & -0.1 & 5,  \\
NGC 0828 & 12 & \dots & 3.11 & 2.83 & 3.11 & 1.46 & 0.09 & 0.11 & 4,  \\
NGC 1068 & 7.3 & \dots & 4.46 & 3.75 & 4.46 & 3.8 & 1.92 & 2.21 & BG066,  \\
NGC 1144 & 10 & \dots & 2.82 & 2.63 & 2.82 & 1.59 & -0.08 & 0.44 & AC205,  \\
NGC 1365 & 20 & \dots & 3.45 & 3.06 & 3.45 & 2.7 & 0.55 & 0.23 & AC348,  \\
NGC 1614 & 2.2 & \dots & 4.3 & 3.64 & 4.3 & 3.22 & 1.96 & 1.71 & AL503,  \\
NGC 2146 & 27 & \dots & 3.02 & 2.76 & 3.02 & 2.21 & 0.44 & 0.42 & 3,  \\
NGC 2369 & 28 & \dots & 2.33 & 2.3 & 2.33 & 1.53 & -0.43 & -0.48 & 10,  \\
NGC 2388 & 5.5 & \dots & 3.41 & 3.03 & 3.41 & 2.45 & 0.87 & 0.58 & 11,  \\
NGC 2623 & 0.46 & \dots & 5.39 & 4.38 & 5.39 & 4.49 & 3.07 & 2.91 & 7,  \\
NGC 3034 & 43 & \dots & 3.46 & 3.07 & 3.46 & 2.44 & 1.05 & 0.87 & AW624+AS250,  \\
NGC 3110 & 20 & \dots & 2.16 & 2.18 & 2.16 & 1.11 & -0.38 & -0.24 & 5,  \\
NGC 3256 & 12 & \dots & 3.49 & 3.09 & 3.49 & 2.56 & 0.95 & 0.91 & AS412+AR531,  \\
NGC 3690 & 27 & \dots & 2.41 & 2.35 & 2.41 & 1.63 & 0.3 & 0.21 & AY102,  \\
NGC 5135 & 6.5+0.7 & \dots & 3.48 & 3.08 & 3.48 & 2.75 & 0.76 & 0.91 & 8,  \\
NGC 5653 & 19 & \dots & 2.27 & 2.25 & 2.27 & 1.19 & -0.32 & -0.44 & 5,  \\
NGC 6240 & 8.6 & \dots & 3.09 & 2.81 & 3.09 & 2.67 & 0.57 & 0.98 & 4,  \\
NGC 6701 & 22 & \dots & 2.38 & 2.33 & 2.38 & 1.32 & -0.49 & -0.48 & 4,  \\
NGC 6921 & 4.1 & \dots & 3.8 & 3.3 & 3.8 & 3.04 & 0.97 & 0.46 & AC264,  \\
NGC 7130 & 0.6 & \dots & 5.51 & 4.46 & 5.51 & 4.71 & 2.79 & 2.96 & AK394,  \\
NGC 7469 & 3.3 & \dots & 3.91 & 3.37 & 3.91 & 3.03 & 1.56 & 1.47 & AL490,  \\
NGC 7552 & 7.2+6.1 & \dots & 3.5 & 3.09 & 3.5 & 2.64 & 1.09 & 0.74 & AS721,  \\
NGC 7591 & 0.94 & \dots & 4.71 & 3.91 & 4.71 & 3.59 & 2.07 & 1.92 & AC685,  \\
NGC 7771 & 16.97 & \dots & 2.59 & 2.48 & 2.59 & 2.05 & 0.01 & -0.06 & 5,  \\
UGC 1845 & 3.1 & \dots & 3.8 & 3.29 & 3.8 & 2.88 & 1.15 & 1.07 & AC301+AT149,  \\
UGC 5101 & 0.41 & \dots & 5.12 & 4.2 & 5.12 & 4.84 & 2.93 & 3.26 & AA095+AS402,  \\
VII Zw 31 & 3.6 & \dots & 3.65 & 3.19 & 3.65 & 2.66 & 0.76 & 0.76 & 4,  \\
ZW 049.057 & 0.54 & \dots & 5.1 & 4.18 & 5.1 & 4.7 & 2.92 & 2.5 & 5,  \\
\enddata

\tablerefs{
	\mbox{{\bf Radio map:} [1] \citet{condon1987}; [2] \citet{buczilowski1987}; [3] \citet{braun2007};
	[4] \citet{condon1996}; [5] \citet{condon1990}; [6] \citet{irwin1999};}
	\mbox{[7] \citet{condon1991}; [8] \citet{ulvestad1989}; [9] \citet{condon2002}; [10] \citet{mauch2008};
	[11] NRAO Image Archive; [12] \citet{becker2003}. {\bf $\Sigma_{\rm H\,{\scriptsize I}}$:}}
	\mbox{[13] \citet{walter2008}; [14] \citet{chung-A2009}; [15] \citet{vanderhulst2002}; 
	[16] \citet{hunter2012}; [17] \citet{martin1998}; [18] \citet{cayatte1994}; [19] \citet{rhee1996};}
        \mbox{[20] \citet{chamaraux1986}; [21] \citet{giovanardi1983}; [22] \citet{fouque1984}; [23] \citet{bottinelli1970};
	[24] \citet{whiteoak1977}; [25] \citet{laine1998}; }
	\mbox{[26] \citet{ott2001}; [27] \citet{stanford1990}; [28] \citet{ondrechen1989};  [29] \citet{saikia1990}; 
	[30] \citet{irwin1991}; [31] \citet{noordermeer2005}; [32] \citet{hurt1996};} 
	\mbox{[33] \citet{ryanweber2003}; [34] \citet{freeland2009}; [35] \citet{helou1981};
	[36] \citet{liszt1995}; [37] \citet{kobulnicky1995}; [38] \citet{oosterloo2010};}
	\mbox{[39] \citet{gallagher1975}; [40] \citet{li1994}.}
 }

 \tablecomments{ \mbox{a - the molecular gas mass (or total gas mass) is calculated using a fixed 
	 CO-to-${\rm H_2}$ conversion factor value of 4.6 ${\rm M_\odot~(K~km~s^{-1}~pc^2)^{-1}}$;
           b - the molecular gas} 
        \mbox{mass is calculated using continuously-varying $\alpha_{\rm CO}$. The (J)VLA observation 
	project codes were given for the radio images reduced in this work. 
        Two project codes were given} 
\mbox{in the cases that the data sets from two different (J)VLA configurations 
were combined. For interacting galaxies which clearly show two independent components, such as}
\mbox{Arp\,055 and Arp\,220, we gave the radio sizes for both components.}
 }

\end{deluxetable*}

\clearpage

%%%%%%%%%%%%%%%%%%%%%%%%%%%%%%%%%
%appendix
%%%%%%%%%%%%%%%%%%%%%%%%%%%%%%%%%
\appendix

%%%%%%%%%%%Appendix 1. Calculating Gas Mass and SFR
\section{A. Calculating Gas Mass and SFR} \label{appendix:gas mass and SFR}
\subsection{A.1 Atomic Gas Mass} \label{appendix:gas mass}
As H\,{\scriptsize I} emission is optically thin for most galaxies and does not suffer from extinction by interstellar dust, emission of H\,{\scriptsize I} can be converted directly into H\,{\scriptsize I} mass. The H\,{\scriptsize I} masses have been calculated from the H\,{\scriptsize I} fluxes using:
\footnotesize
\begin{equation}
	M({\rm H\,{\scriptsize I}})~{\rm [M_\odot]} = 2.36 \times 10^5 \times D_{L}^2 \times F({\rm H\,{\scriptsize I}})
\end{equation}
\normalsize
where $D_{L}$ is the distance to the source in Mpc and F(H\,{\scriptsize I}) is the H\,{\scriptsize I} flux in units of Jy km $s^{-1}$. Given the fact that most of our objects are seen at low inclinations, we did not attempt to correct for any unknown optical depth effects. We assume that H\,{\scriptsize I} emissions in all galaxies are optically thin. If there were significant self-absorption, this would cause the derived H\,{\scriptsize I} masses to be underestimated.

\subsection{A.2 Molecular Gas Mass}
The CO line luminosity $L^\prime_{CO}$ was calculated from the formula: 
\footnotesize
\begin{equation}
	L^\prime_{CO} = 3.25 \times 10^7 \times S_{\rm CO}~\Delta \upsilon \times \nu_{obs}^{-2} \times D_{L}^2 
\end{equation}
\normalsize
where $L^\prime_{CO}$ is in ${\rm K~km~s^{-1} ~pc^2}$, 
${S_{\rm CO}~\Delta \upsilon}$ in Jy km ${\rm s^{-1}}$, distance $D_{L}$ in Mpc 
and observation frequency $\nu_{obs}$ of CO (1-0) for nearby galaxies is 115.27 GHz. 
The molecular gas mass can be derived from $L^\prime_{CO}$
with an empirical derivation of the CO-${\rm H_2}$ conversion factor ($\alpha_{CO}$) 
which assumes a linear conversion relation:
$M({\rm H_2+He})~({\rm M_\odot})= \alpha_{CO} \times L^\prime_{CO}~{\rm (K~km~s^{-1})}$.
The Galactic conversion factor value is $\alpha_{CO}
={\rm 4.6~M_\odot~(K~km~s^{-1} ~pc^2)^{-1}}$ \citep{young1991}.
A correction factor of 1.36 was applied to include helium. 

\subsection{A.3 Dense Molecular Gas Mass}
The line luminosity of dense molecular gas $L^\prime_{\rm HCN}$ is then determined with the formula (GS04b):
\footnotesize
\begin{equation} 
	L^\prime_{\rm HCN} ~{\rm (K~km~s^{-1}~pc^2)^{-1}}= 3.25 ~\times~10^7~\times~S_{\rm HCN}~\Delta \upsilon \times \nu_{obs}^{-2} \times D_{L}^2,
\end{equation}
\normalsize
where ${S_{\rm HCN}~\Delta \upsilon}$ in Jy km ${\rm s^{-1}}$, 
distance $D_{L}$ in Mpc and observation frequency $\nu_{obs}$ of HCN (1-0) for local galaxies is 88.6 GHz. 
The dense molecular gas mass $M_{\rm dense}$ is obtained from $L^\prime_{\rm HCN}$ using the formula given by GS04a:
\footnotesize
\begin{equation}
M_{\rm dense}~({\rm M_\odot}) = 10 \times L^\prime_{\rm {HCN}} ~{\rm (K~km~s^{-1}~pc^2)^{-1}}
\end{equation}
\normalsize
which assumed a conversion factor of $10~{\rm M_\odot ~(K ~km~s^{-1}~pc^2)^{-1}}$  between ${\rm L^\prime_{\rm HCN}}$ and dense molecular gas mass.

\subsection{A.4 The SFRs} \label{appendix:SFRs}
The SFRs calibration of \citet{bell2003} which takes into account the effect of the reduced efficiency of IR 
\begin{wrapfigure}{}{0.41\textwidth}
\includegraphics[scale=0.40]{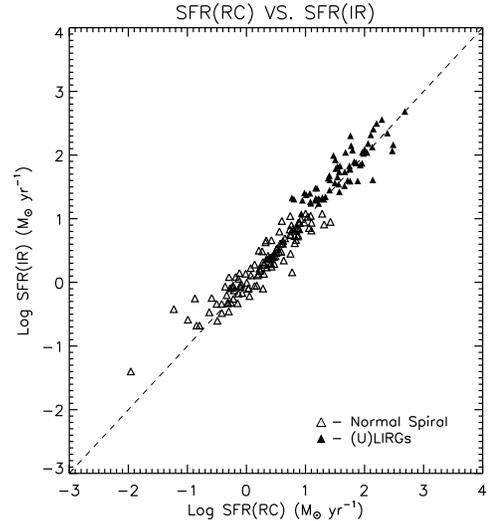}

\caption{\scriptsize The distributions of SFR(IR) and SFR(RC) generally follow a trend denoted 
by the dashed line which is linear, suggesting that the two SFRs gave qualitatively identical results.
}
\label{figure:IR_RC}
\end{wrapfigure}
and radio emission from low-luminosity galaxies, 
along with correcting for old stellar populations,  is
\footnotesize
\begin{equation}
{\rm SFR(IR)}=\left\{
   \begin{array}{c}
   1.57 \times 10^{-10}L_{\rm IR}(1+\sqrt{10^9/L_{\rm IR}}), ~~~L_{\rm IR} > 10^{11}, \\
   1.17 \times 10^{-10}L_{\rm IR}(1+\sqrt{10^9/L_{\rm IR}}), ~~~L_{\rm IR} \leq 10^{11}, \\
   \end{array}
  \right.
  \end{equation}

 \begin{equation}
 {\rm SFR(RC)}=\left\{
   \begin{array}{c}
   5.52 \times 10^{-22} L_{\rm 1.4GHz},~~~~~~~~~~~~~~~~~~~~~~~~~~~~~~~L>L_c,\\
   (5.52 \times 10^{-22} L_{\rm 1.4GHz}) / [0.1+0.9(L_{\rm 1.4GHz}/L_c)^{0.3}],~~~L \leq L_c.  \\
   \end{array}
  \right.
  \end{equation}
 \normalsize

Here, we define SFR(IR) and SFR(RC) as the SFRs derived from total infrared luminosity 
and 1.4 GHz radio continuum luminosity, respectively.
The SFRs is in units of ${\rm M_\odot~yr^{-1}}$, $L_{\rm IR}$ is given in units of ${\rm L_\odot}$,
${\rm L_{1.4GHz}}$ is in ${\rm W~Hz^{-1}}$ and $L_c = 6.4 \times 10^{21}~{\rm W~Hz^{-1}}$,
which is the 1.4 GHz radio luminosity of a $\sim L_\ast$ galaxy.

Fig.\ \ref{figure:IR_RC} shows the comparison of the IR-based SFR(IR) and RC-based SFR(RC).
Overall, the two SFRs gave qualitatively identical results with scatter of only $\sim 0.21$ dex.
It is worth a quick note that the two SFRs of some individual galaxies 
can deviate substantially from each other, 
such as NGC 4394 and IRAS 12243-0036.
This was also noted by \citet{bell2003}, i.e., 
the above calibrations are more based on a galaxy-by-galaxy basis
and some individual galaxies can deviate substantially from the mean behavior. 
The SFRs in our sample spanned a wide range, 
from $\leq 1~{\rm M_\odot~yr^{-1}}$ in some quiescent early type galaxies
to more than $\sim 480~{\rm M_\odot~yr^{-1}}$ in strong starbursts. 

\begin{wrapfigure}{l}{0.41\textwidth}
	\includegraphics[scale=0.40]{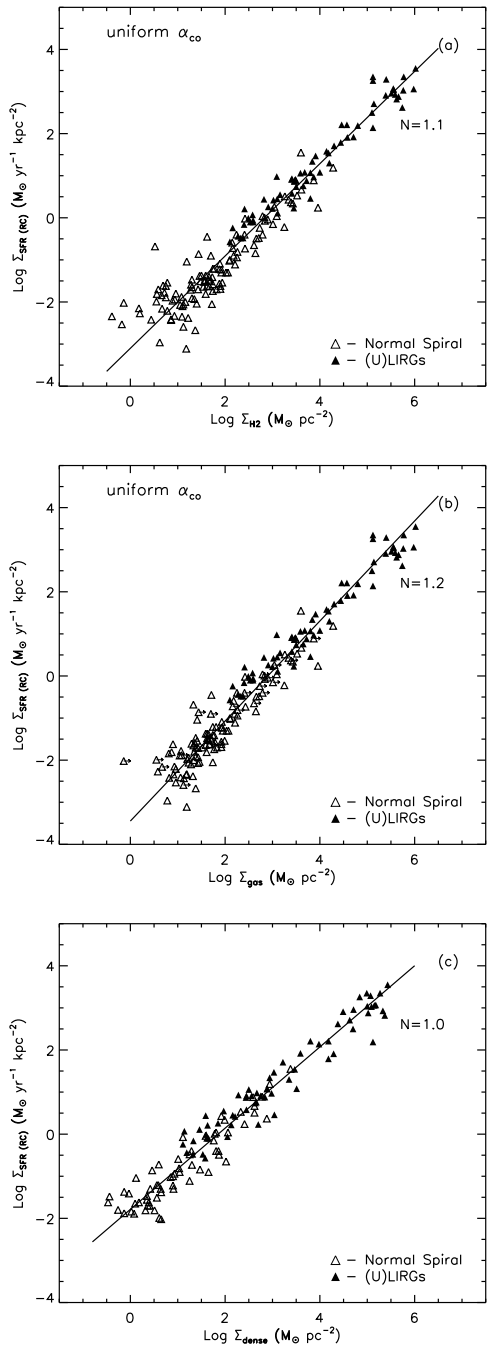} 
	\caption{\scriptsize The SF laws derived by using RC-based SFRs: 
		(a) SFR(RC) vs. ${\rm H_2}$; (b) SFR(RC) vs. total gas; (c) 
		SFR(RC) vs. dense gas. The molecular gas (and total gas) here
	 is deduced by adopting contant Galactic $\alpha_{\rm CO}$. The SF laws derived by
 adopting RC-based SFRs agree well with the results using IR-based SFRs.}
\label{figure:KSL_RC}
\end{wrapfigure}

As an excellent agreement was found between the two SFR estimators,
using one over the other, therefore, did not alter our results of SF laws in any
significant way. Here we show some examples of SF laws derived by adopting RC-based SFRs.
Fig.\ \ref{figure:KSL_RC} presents the relations between the densities of SFR(RC) 
and molecular gas, total gas as well as dense gas.
The derived slopes are $N=1.10 \pm 0.02$, $N=1.19 \pm 0.02$ and $N=0.98 \pm 0.02$, respectively,
agreeing very well with the same relations derived from IR-based SFRs.

%==========================
%%%%%%%%%%%Appendix 2. Radio Size Measurement
\section{B. Radio Size Measurement} \label{appendix:radio size measurement}
Most of our radio images are based on 1.4 GHz (L-band) RC observations.
However, for the compact starbursts whose radio emissions are 
unresolved by the highest resolution ($\sim 1\arcsec.5$) of 
1.4 GHz VLA maps, the higher frequency 
(J)VLA observations were adopted to achieve higher resolutions.  
It is worth noting that the high-resolution maps may miss extended low-brightness 
components, and thus the extended global structure of the source. 
We made every effort to ensure that each galaxy can be resolved
yet little missing flux due to high interferometer resolution. 

The galaxy major/minor axis was determined by fitting 2D Gaussian to each 
radio map using the  $\cal AIPS\/$ task IMFIT. 
In general, we tend to choose the resolved radio maps with the lowest resolution
for each galaxy. 
Around 98\% (178/181) of our galaxies were resolved, among which
$\sim$ 88\% (158/181) were fully resolved with deconvolved sizes larger than beam sizes,  
and $\sim$ 11\% (20/181) galaxies were partially resolved 
(i.e., those with deconvolved sizes larger than 1/2 beam size).
As for the cases where one-component Gaussian model failed,
such as the galaxies with complex morphologies,
we measured their effective diameters, which contains
half of the integrated radio flux, instead.

The radio sizes of our galaxies range between $0\arcsec.2$ to $13\arcmin$,
corresponding to $\sim$ 0.2 kpc to 20 kpc in linear scale.
The typical values for normal spirals were $1-10$ kpc,
while the size of (U)LIRGs were significantly smaller
with typical values being a few kpc or less.
Especially for those extremely compact starbursts, their high surface brightness radio emission 
is restricted at nuclear region within an extent of a few hundreds pc scale. 
%Table 2 illustrates the range of properties of the galaxies, i.e., 

One important caveat on the use of radio size, however,
 is the possibility of an active galactic nucleus (AGN)
contributing to the radio emission.
Luckily, the interferometers with very long baselines (e.g., VLBA and VLBI) 
have extremely high resolution to easily pick out AGN.
Our trick is to assume AGN are point sources and estimate their radio contribution 
through highest angular resolution observations to galaxy radio cores by VLBA/VLBI. 
We found that radio cores contributed only a small fraction ($\leq 10\%$) of 
the total radio flux for most of our Seyfert galaxies.
As for those Seyfert galaxies whose radio core contribute more than $10\%$
of the total RC flux, we subtracted point source equivalent to that value 
in lower resolution VLA map and then fitted a Gaussian to the difference. 
Thus, we took the average of the upper and lower values 
as a approximate estimate of the radio size in these Seyfert galaxies.

%%%%%%%%%%%Appendix 2. IR/HCN to FIR color
\section{C. Correlation with FIR Color C(60/100)} \label{appendix:FIR color}

\begin{wrapfigure}{}{0.41\textwidth}
	\includegraphics[scale=0.40]{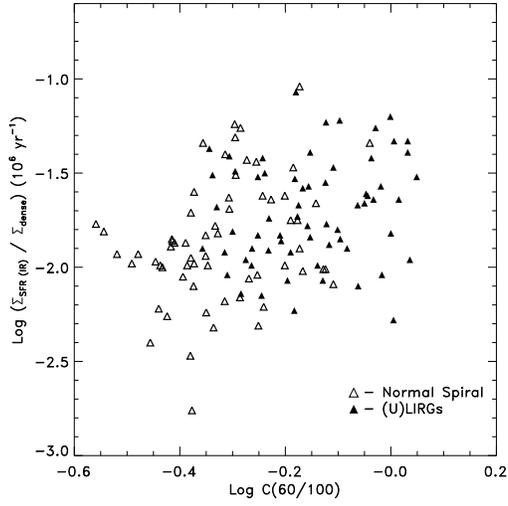} 
	\caption{$\Sigma _ {\rm {SFR}}$/$\Sigma _ {\rm {dense}}$ ratio
	as a function of the FIR color [C(60/100)].}
\label{figure:FIR_color}
\end{wrapfigure}

Fig.\ \ref{figure:FIR_color} presents $\Sigma _ {\rm {SFR}}$/$\Sigma _ {\rm {dense}}$ versus C(60/100) for HCN sample.
The $\Sigma _ {\rm {SFR}}$/$\Sigma _ {\rm {dense}}$ shows a weak dependence
on the C(60/100), or dust temperature $T_{\rm dust}$ since C(60/100) is a good
estimator of $T_{\rm dust}$ \citep{chanial2007,diaz-santos2010}. 
The $\Sigma _ {\rm {SFR}}$/$\Sigma _ {\rm {dense}}$ ratio gets slightly higher
for warmer dust temperature.

\begin{wrapfigure}{}{0.41\textwidth}
	\includegraphics[scale=0.40]{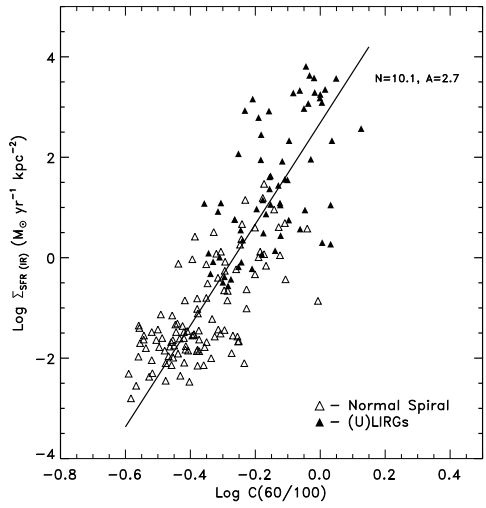} 
	\caption{$\Sigma _ {\rm {SFR}}$ as a function of the FIR color [C(60/100)].}
\label{figure:FIR_color2}
\end{wrapfigure}

We found a rather strong FIR color denpence of $\Sigma _ {\rm {SFR}}$.
Fig.\ \ref{figure:FIR_color2} plots our IR-based SFR versus C(60/100) for the 
175 galaxies with $C(60/100) > 0.25$. 
Fitting all the points in the figure derives a relation:
${\rm \log \Sigma_{SFR} = (10.09 \pm 0.46) \log C(60/100) + (2.68 \pm 0.17)}$.
And this fitted relationship was used in Lu et al. (ApJL, submiited) 
in roughly estimating SFR surface density from FIR color. 

\clearpage

\bibliographystyle{apj}
\bibliography{references}
\clearpage
\end{document}